\documentclass[11pt]{article}

\usepackage[a4paper,margin=1in]{geometry}
\usepackage{graphicx}
\usepackage{amsmath,amssymb}
\usepackage{color}
\usepackage{hyperref}
\usepackage{url}

\usepackage[utf8]{inputenc}

\begin{document}
	
	\title{\large\bf Dynamical Evolution and Graceful Exit in Quartic Warm Inflation}
	
	\author{
		Bhargabi Saha\footnote{\tt bhargabi@iitg.ac.in}\ \ and
		Malay K. Nandy\footnote{\tt mknandy@iitg.ac.in {\rm (Corresponding Author)}} \\
		{\em Department of Physics, Indian Institute of Technology Guwahati, India}
	}
	
	\date{\small (November 19, 2025)}
	\maketitle
	
	\begin{abstract}
		In this study, we investigate the full nonlinear dynamics of warm inflation driven by the quartic inflaton potential, avoiding any simplifying approximations.  The thermal backreaction is incorporated through a dissipation coefficient that depends linearly on the temperature, and the model parameters are chosen to remain consistent with Planck observational constraints. By numerically integrating the complete set of three coupled, nonlinear differential equations that describe the evolution of the inflaton field, radiation energy density, and background expansion, we obtain an exact description of the system’s dynamics. 
		Our results reveal that, while transition to radiation domination is suppressed in the weak regime, the strong dissipative regime leads to a smooth and natural transition to a hot, radiation-dominated Universe, thereby confirming graceful-exit within warm inflation in the quartic scenario. The reheating temperature is extracted directly from the nonlinear evolution of warm inflation, yielding a temperature of approximately $\sim 10^{13}$ GeV at the end of inflation, which cools to about $\sim 10^{12}$ GeV near radiation–inflaton equality, whence the Universe transitions into a radiation-dominated era.
	\end{abstract}
	

\section{Introduction}
\label{sec1}

The scenario of {\em warm inflation} was first proposed by Berera in the mid-1990s \cite{Berera1995,Berera_Fluc_1995,Berera_1996,Berera_1997,BERERA2000,Berera_2005} as a conceptual alternative to the conventional {\em cold inflationary paradigm} 
\cite{Sato:1980yn,Linde-inflation1982,linde1983chaotic,linde1984inflationary,guth1984inflationary,Albrecht1982A,albrecht1982reheating,HAWKING198235}. In this framework, the inflaton is not an isolated field but interacts dynamically with a surrounding thermal bath. Such interactions introduce a {\em dissipative} (friction) term in the inflaton’s equation of motion, arising naturally within nonequilibrium quantum field theory \cite{Hosoya1984}. The resulting thermal damping can, under suitable conditions, compete with or even dominate over Hubble friction, thereby moderating the inflaton’s evolution \cite{MOSS1985,YOKOYAMA1988,LIDDLE19895}. Consequently, thermal rather than quantum fluctuations become the primary source of primordial density perturbations \cite{MOSS1985}.

Specifically, the {\em cold inflation scenario} involves a rapid exponential expansion that is followed by a distinct reheating stage, during which the inflaton decays into radiation \cite{kofman1994reheating,kofman1996origin,albrecht1982reheating,abbott1982particle,allahverdi2010reheating,amin2015nonperturbative}. Warm inflation, on the other hand, incorporates dissipative particle production during inflation itself \cite{Bastero-Gil_2013,Zhang_2009}, making a separate reheating phase unnecessary. This continuous energy exchange has the possibility of enabling a smooth transition from inflation to the radiation-dominated epoch. Such behavior may offer elegant resolution to issues that challenge cold inflation, including the graceful-exit problem \cite{LINDE1990} and the generation of unwanted relics \cite{NAGANO1998}. Numerous theoretical realizations have since been developed, in which quantum field interactions between the inflaton and additional light or heavy fields cause dissipation in the early Universe. Recent analyses demonstrate that warm inflation models can remain consistent with Planck constraints \cite{Kamali_2015,Setare2015,Kamali2016,Visinelli_2016,Panotopoulos2015,Kamali2018,Jawad2017,Jawad-2017,Herrera2015,AlHallak2023}, reinforcing their status as viable descriptions of the inflationary epoch and the origin of large-scale structure. Moreover, by lowering the predicted tensor-to-scalar ratio, warm inflation can render some previously excluded potentials compatible with current CMB observations \cite{Benetti2017}.

A key characteristic of warm inflation is the {\em dissipative ratio}, $R = \Gamma / (3H)$, which measures the relative strength of dissipation to Hubble friction \cite{Bastero-Gil_2013,Zhang_2009}. When $R \ll 1$, the system behaves similarly to standard cold inflation, while for $R \gg 1$, energy transfer from the inflaton to radiation becomes efficient, enabling inflation to proceed even at lower potential energies with interaction between the inflaton and other fields \cite{Berera-2001,Berera_2023}.

In addition to providing a naturally sustained mechanism for reheating, warm inflation offers a promising interface with high-energy physics frameworks, such as supersymmetric and string-theoretic models \cite{Bastero-Gil-2019,basterogil2009}. The interplay between thermal noise and dissipative dynamics can lead to distinctive signatures in key observables, including the scalar spectral index $n_s$ and the tensor-to-scalar ratio $r$ \cite{Herrera2018}, allowing for direct confrontation with precision cosmological data.

Extensions of the warm inflation concept have been used to connect the early and late accelerating phases of the Universe. One such unified model \cite{DIMOPOULOS2019} employs a scalar potential that behaves as a quartic chaotic form for $\phi < 0$ and transitions to an inverse quartic power-law quintessence for $\phi > 0$ within a weak dissipative regime. During inflation, thermal fluctuations amplify scalar perturbations while maintaining a residual radiation bath that naturally reheats the Universe. At later times, the same scalar field can drive dark-energy-like acceleration, with its potential being tightly constrained.

An alternative formulation interprets warm inflation through the thermodynamics of open systems \cite{HARKO2020}, treating the decay of the inflaton into radiation as an irreversible process that introduces a creation pressure into the energy-momentum tensor. In such models, particle production peaks near the end of inflation, and the resulting predictions, especially for Higgs-type potentials, are in good agreement with Planck constraints.

The $\alpha$-attractor E-model has recently been analyzed within the framework of warm inflation \cite{Saha-2025}. With a dissipation coefficient linear in temperature, the study derived analytical expressions for key observables, such as the scalar spectral index $n_s$ and tensor-to-scalar ratio $r$, under slow-roll conditions. The results confirmed that the model remains fully consistent with Planck 2018 constraints across both weak and strong dissipative regimes.

In this work, we focus on the consequences of the nonlinear dissipative dynamics of the inflaton field, which strongly couples to the ambient radiation field, a fundamental feature of warm inflation. To account for the {\em nonlinear} nature in a simplified yet explicit manner, we adopt the quartic inflaton potential, which has been shown to be consistent with cosmological observations \cite{Panotopoulos2015}. Additionally, we consider a dissipation coefficient proportional to temperature. To accurately trace the evolution, we solve the full set of coupled, nonlinear dynamical equations without resorting to approximations. Our analysis reveals that, while insufficient dissipation prevents a transition to radiation domination in the weak regime, the strong dissipative regime leads to a smooth and natural transition to a hot, radiation-dominated Universe, thereby confirming graceful-exit within warm inflation in the quartic scenario.

The remainder of the paper is organized as follows. In Section \ref{sec2}, we present the nonlinear dynamics of warm inflation, modeling the inflaton energy density using a quartic potential and deriving the rate equations for both the inflaton and radiation energy densities, with a dissipative coefficient that scales with temperature. In Section \ref{basics}, we define the weak and strong dissipation regimes of warm inflation and establish the relevant initial conditions. The dynamical equations derived in Section \ref{basics} are then formulated as a system of three coupled nonlinear differential equations in Section \ref{sec4}.  We numerically solve these equations in Section \ref{sec4} and analyze the time evolution of key quantities. Finally, in Section \ref{Discussion}, we conclude with a discussion that highlights various features of warm inflation originating from the inherent nonlinearity in the context of the quartic model.

\section{Dynamics of warm inflation}
\label{sec2}

In order to focus upon the dynamics of warm inflation, we shall consider the quartic potential for the inflaton field $\phi$, given by
\begin{equation}
\label{potential}
V(\phi)=\frac{1}{4}\lambda\phi^4,
\end{equation}
which has been rescued as a viable model of warm inflation.
In the presence of radiation energy density $\rho$, the dynamics is governed by the Friedman equation
\begin{equation}
\label{fried1}
H^2=\frac{1}{3M_P^2} \left(u+\rho\right),
\end{equation}
where $M_P=\frac{1}{\sqrt{8\pi G}}$ is the reduced Planck mass and $u$ is the inflaton energy density, given by
\begin{equation}
\label{inflaton}
u = \frac{1}{2}\dot \phi^2+ V(\phi),
\end{equation}
with the pressure of the inflaton field
\begin{equation}
\label{pressure}
P = \frac{1}{2}\dot \phi^2- V(\phi).
\end{equation}

In the warm inflation scenario, time evolutions of the inflaton energy density $u$ and the radiation energy density $\rho$ can be modelled as
\begin{equation}
\label{KG}
\dot u+3H(u+P)=-\Gamma\dot\phi^2
\end{equation}
and
\begin{equation}
\label{reheat}
\dot\rho+3H(\rho+p)=\Gamma\dot\phi^2,
\end{equation}
where $\Gamma$ is the dissipative coefficient, and the radiation equation of state is given by $p=\frac{1}{3}\rho$.

Warm inflation fundamentally relies on the existence of thermalized radiation during the inflationary era. The energy density of this radiation is given by the relation:
\begin{equation}
\label{radiation}
\rho=C_*T^4,
\end{equation}
 $T$ is the temperature and $C_*=\frac{\pi^2}{30}g_*$, with $g_*= 228.75$ denoting the effective number of relativistic degrees of freedom. This value of $g_*$ accounts for the contributions from 28 bosons and 90 fermions in the Standard Model, in addition to four higgsinos and the swallowed scalar Higgs in the minimal supersymmetric model.

Using \eqref{inflaton}, equation \eqref{KG} gives the dynamics of the inflaton field as
\begin{equation}\label{KG1}
\ddot\phi+3H\dot\phi+V'(\phi)=-\Gamma\dot\phi.
\end{equation}

Upon neglecting $\ddot\phi$ in the slow roll approximation, we have
\begin{equation}\label{KG1a}
3H\dot\phi+V'(\phi)=-\Gamma\dot\phi.
\end{equation}

We further neglect $\dot\rho$ in equation \eqref{reheat} as required in the warm inflation scenario so that the decrement in radiation energy density due to expansion of the Universe is replenished by decay of the inflaton field \cite{Berera_Fluc_1995}. Consequently, the radiation temperature given by \eqref{radiation} can be obtained as
\begin{equation}
\label{temperature}
T=\left(\frac{\Gamma V'^2}{36C_*H^3(1+R)^2}\right)^{1/4},
\end{equation}
where $R$ is the dissipation ratio, defined as
\begin{equation}
\label{R}
R = \frac{\Gamma}{3H}.
\end{equation}

Slow roll parameters in warm inflation may be defined as \cite{Moss_2008,Hall_2004}
\begin{eqnarray}
\label{slow-roll}
\epsilon=\frac{M_P^2}{2}\left(\frac{V^\prime}{V}\right)^2, \qquad \eta=M_P^2\left(\frac{V''}{V}\right),\nonumber \\
\beta=M_P^2\left(\frac{\Gamma^\prime V^\prime}{\Gamma V}\right), \qquad \sigma=M_P^2\left(\frac{V^\prime}{\phi V}\right),
\end{eqnarray}
with the slow roll conditions
\begin{eqnarray}
\label{condition}
\epsilon \ll 1+R, \qquad \eta \ll 1+R, \nonumber \\
\beta \ll 1+R, \qquad \sigma \ll 1+R.
\end{eqnarray}

In the framework of warm inflation, thermal fluctuations take precedence over quantum fluctuations \cite{Berera_Fluc_1995}, and the amplitude of the power spectrum for curvature perturbations is described by \cite{Berera_1996,Hall_2004,Berera_2009,BASTERO-2009} as:
\begin{equation}
\label{Power}
{A_R }^{1/2}\simeq \left(\frac{3H^3}{2\pi V^\prime}\right)\left(1+R\right)^{5/4}\left(\frac{T}{H}\right)^{1/2}.
\end{equation}

The observed amplitude is typically approximated as ${A_R}^{1/2}\approx 10^{-5}$ \cite{Planck-2013,Planck-2015}. For the purposes of our numerical calculations, we will adopt this value for $A_R$, neglecting the potential variability due to experimental uncertainties.

\begin{figure*}
\centering
\includegraphics[width=0.8\linewidth]{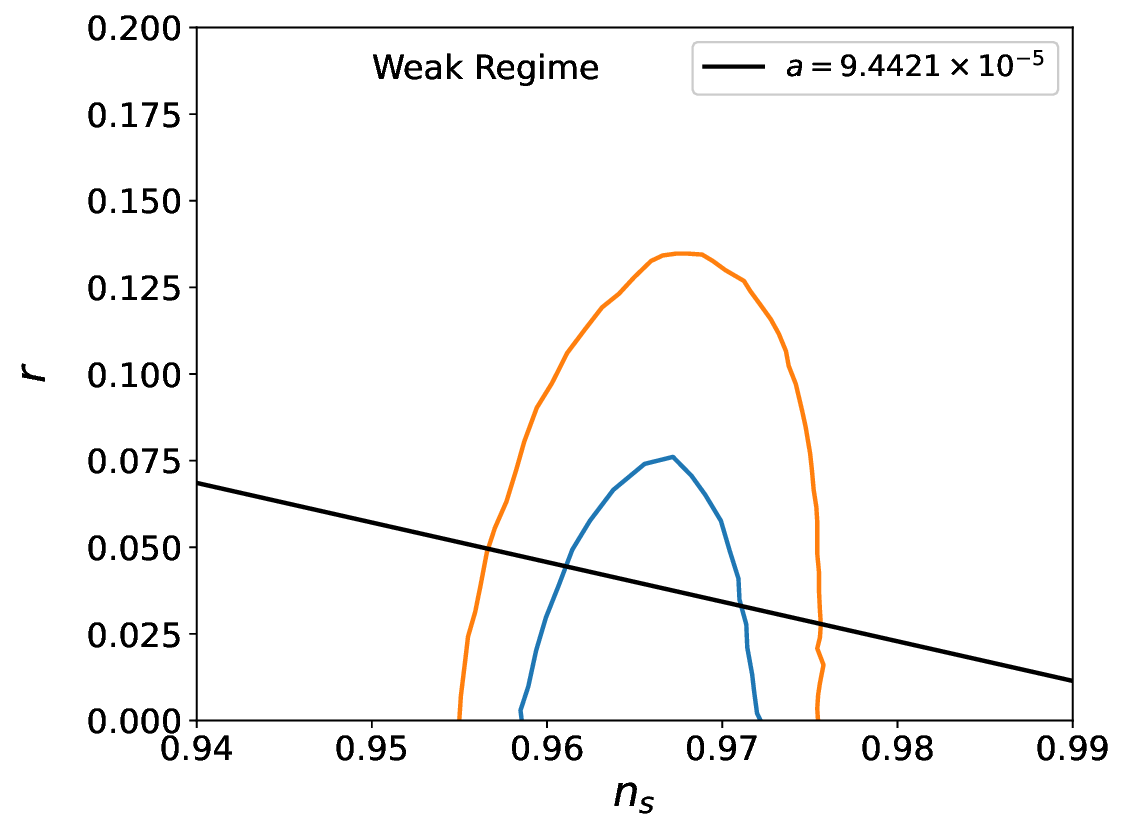}
\caption{Tensor-to-scalar ratio $r$ versus scalar spectral index $n_s$ in the weak dissipative regime for $a=9.4421\times 10^{-5}$. The straight line represents equation \ref{comb1}.}
\label{fig-a}
\end{figure*}

\begin{figure*}
\centering
\includegraphics[width=0.8\linewidth]{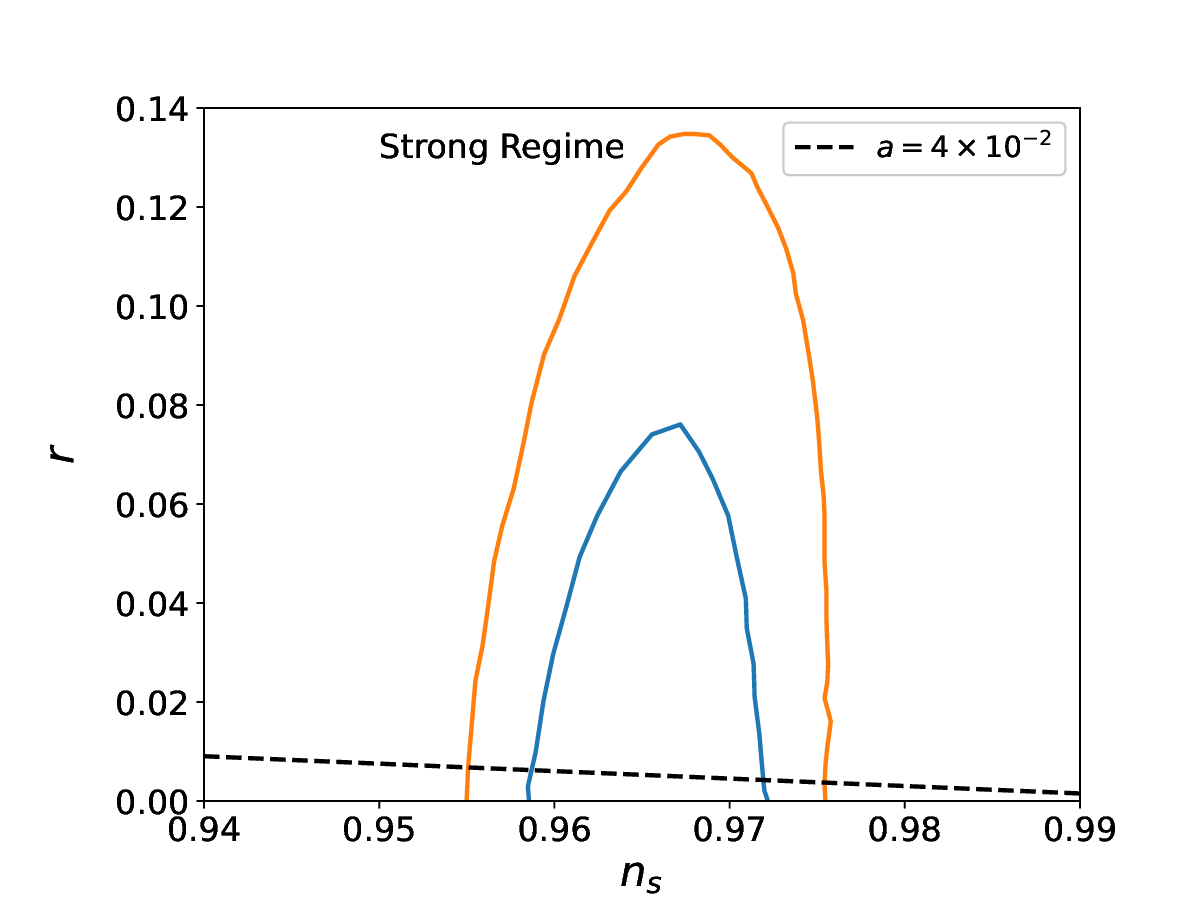}
\caption{Tensor-to-scalar ratio $r$ versus scalar spectral index $n_s$ in the strong dissipative regime for $a=4\times 10^{-2}$. The straight line represents equation \ref{comb2}.}
\label{fig-b}
\end{figure*}

\begin{figure*}
\centering
\includegraphics[width=0.8\linewidth]{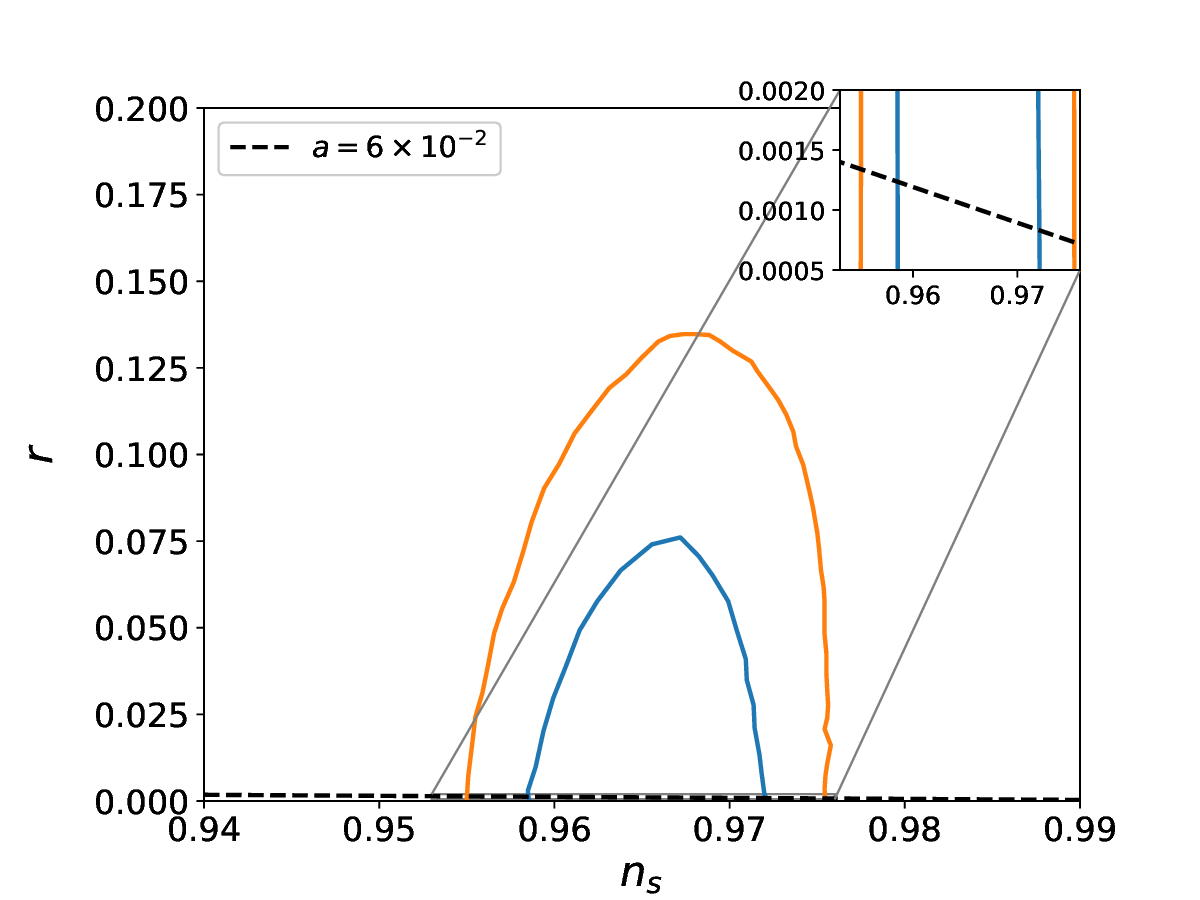}
\caption{Tensor-to-scalar ratio $r$ versus scalar spectral index $n_s$ in the strong dissipative regime for $a=6\times 10^{-2}$. The straight line represents equation \ref{comb2}.}
\label{fig-c}
\end{figure*}

\begin{figure*}
\centering
\includegraphics[width=0.8\linewidth]{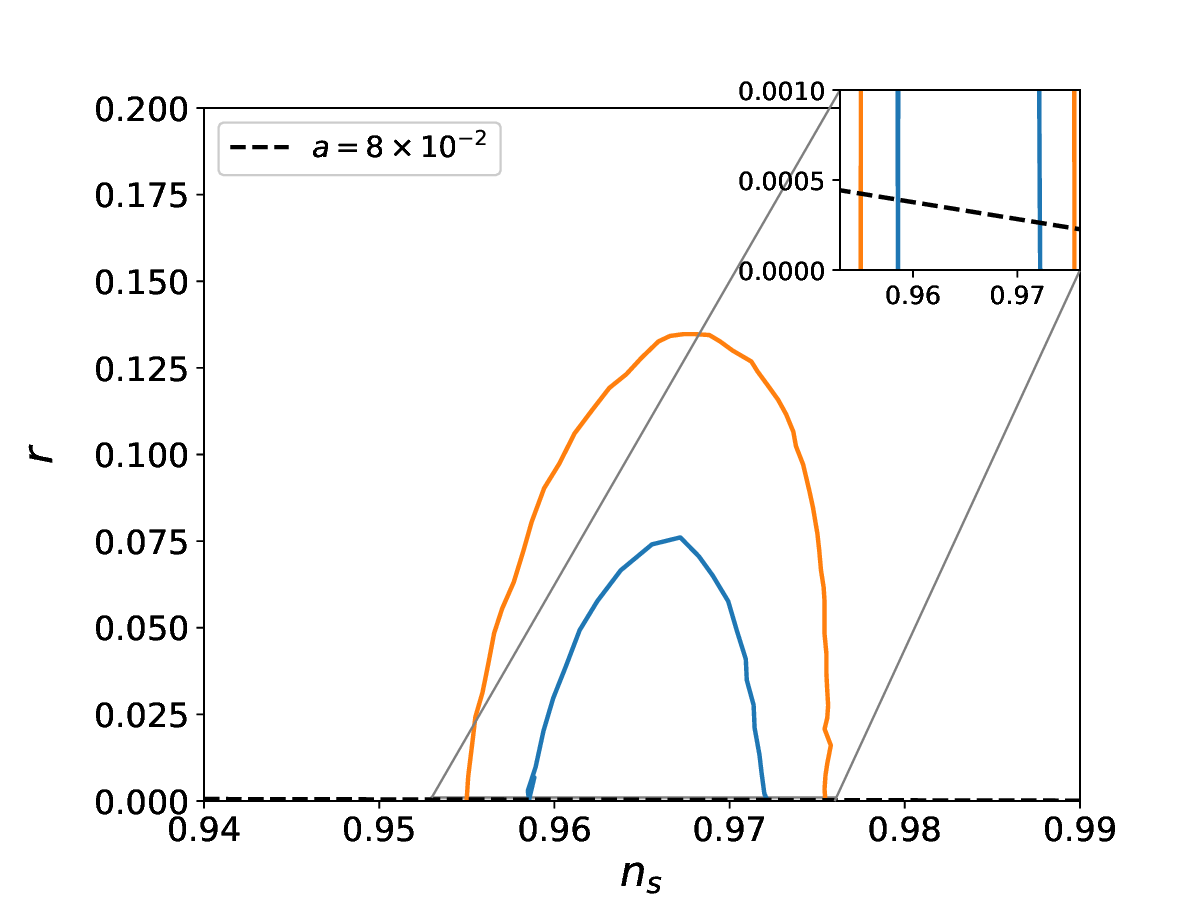}
\caption{Tensor-to-scalar ratio $r$ versus scalar spectral index $n_s$ in the strong dissipative regime for $a=8\times 10^{-2}$. The straight line represents equation \ref{comb2}.}
\label{fig-d}
\end{figure*}

\begin{figure*}
\centering
\includegraphics[width=0.8\linewidth]{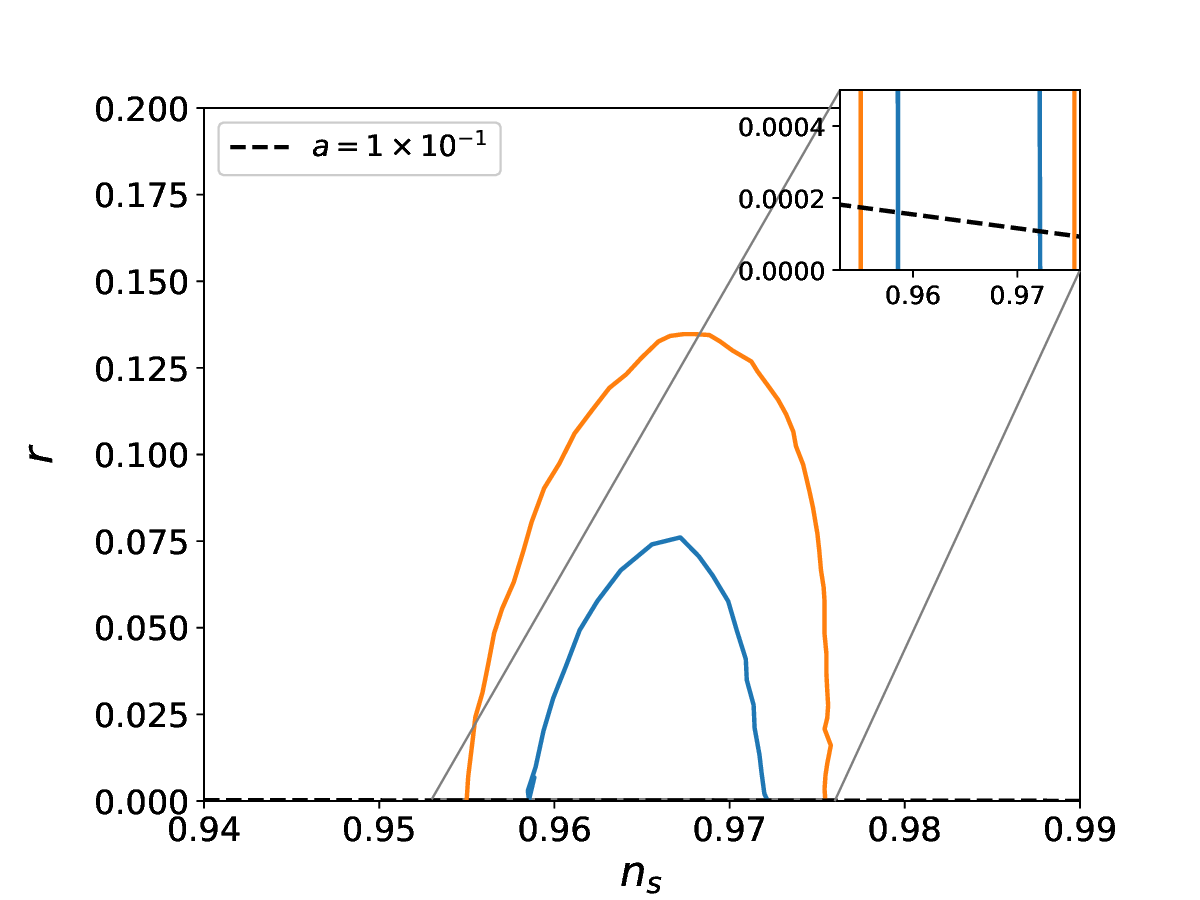}
\caption{Tensor-to-scalar ratio $r$ versus scalar spectral index $n_s$ in the strong dissipative regime for $a=10\times 10^{-2}$. The straight line represents equation \ref{comb2}.}
\label{fig-e}
\end{figure*}

The expressions for the scalar spectral index $n_s$ and the tensor-to-scalar ratio $r$ are given by \cite{Hall_2004,Berera_2009,BASTERO-2009}
\begin{eqnarray}
n_s \simeq & \displaystyle 1+\frac{1}{1+R} \left[ -\left(2-\frac{5R}{1+7R}\right)\epsilon\right. \nonumber\\
& \displaystyle \left. -\left(\frac{3R}{1+7R}\right)\eta + \left(2+\frac{4R}{1+7R}\right)\sigma \right]
\label{ns1}
\end{eqnarray}
and
\begin{equation}
r \simeq  \frac{H}{T} \frac{16\epsilon}{(1+R)^{5/2}}.
\label{r1}
\end{equation}
We shall use these relations to illustrate the validity of the parameters involved within the bounds of the observed Planck data \cite{Planck-2013,Planck-2015}.

\section{Regimes of Warm Inflation: Weak and strong Dissipation}
\label{basics}

In order to investigate the weak and strong regimes of warm inflation, we model the thermal backreaction using a dissipation coefficient given by
\begin{equation}
\label{dissipation}
\Gamma=aT,
\end{equation}
where $a$ is a parameter of the model. This approach, often referred to as ``warm little inflation,'' captures the key characteristics of warm inflation \cite{Bastero_2016}.

\begin{figure*}
\centering
\includegraphics[width=0.8\linewidth]{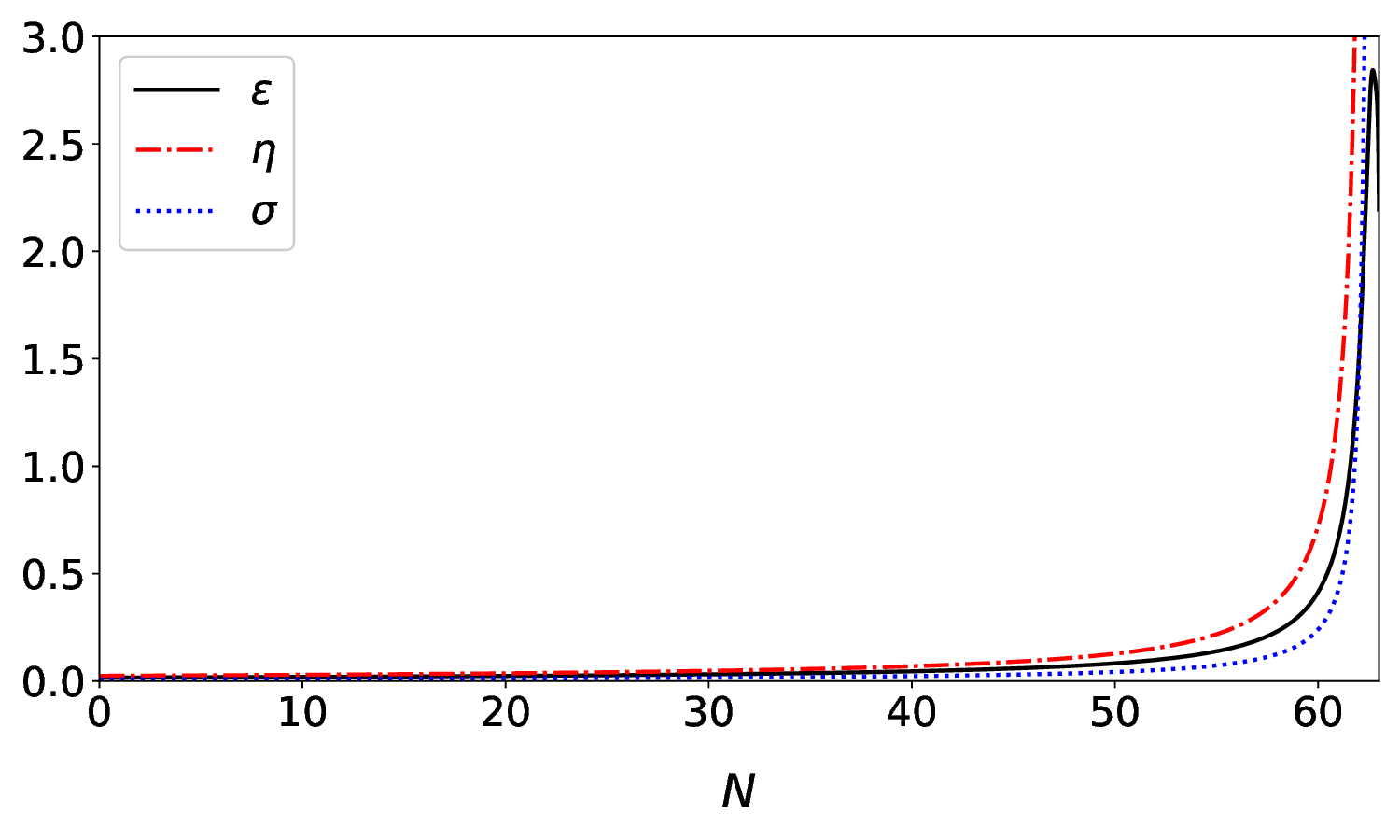}
\caption{Evolution of the slow roll parameters $\epsilon$, $\eta$ and  $\sigma$ as a function of the e-fold number $N$ in the weak dissipative regime. The end of inflation is marked by $\eta=1$, corresponding to $N=61.5986$.}
\label{fig-1}
\end{figure*}

\subsection{The weak dissipative regime}
\label{weak-reg}
The dissipative coefficient is insignificant with respect to the Hubble expansion in the weak regime, $\Gamma\ll 3H$, so that equation \eqref{temperature} reduces to
\begin{equation}
\label{T-weak}
T=\left(\frac{aV^{\prime 2}}{36C_*H^3}\right)^{1/3}.
\end{equation}

Now, with the slow roll condition $\frac{1}{2}\dot\phi^2\ll V(\phi)$, and on the assumption $\rho \ll u$, equation \eqref{fried1} and \eqref{inflaton} lead to
\begin{equation}
\label{fried1a}
H=\left(\frac{V}{3M_P^2}\right)^{1/2}.
\end{equation}

In this regime, upon neglecting $\Gamma$ with respect to $3H$, equation \eqref{KG1a} gives
\begin{equation}
\label{phidot}
\dot\phi=-\frac{V'(\phi)}{3H}=-\frac{\lambda\phi^3}{3H}.
\end{equation}

\begin{figure*}
\centering
\includegraphics[width=0.8\linewidth]{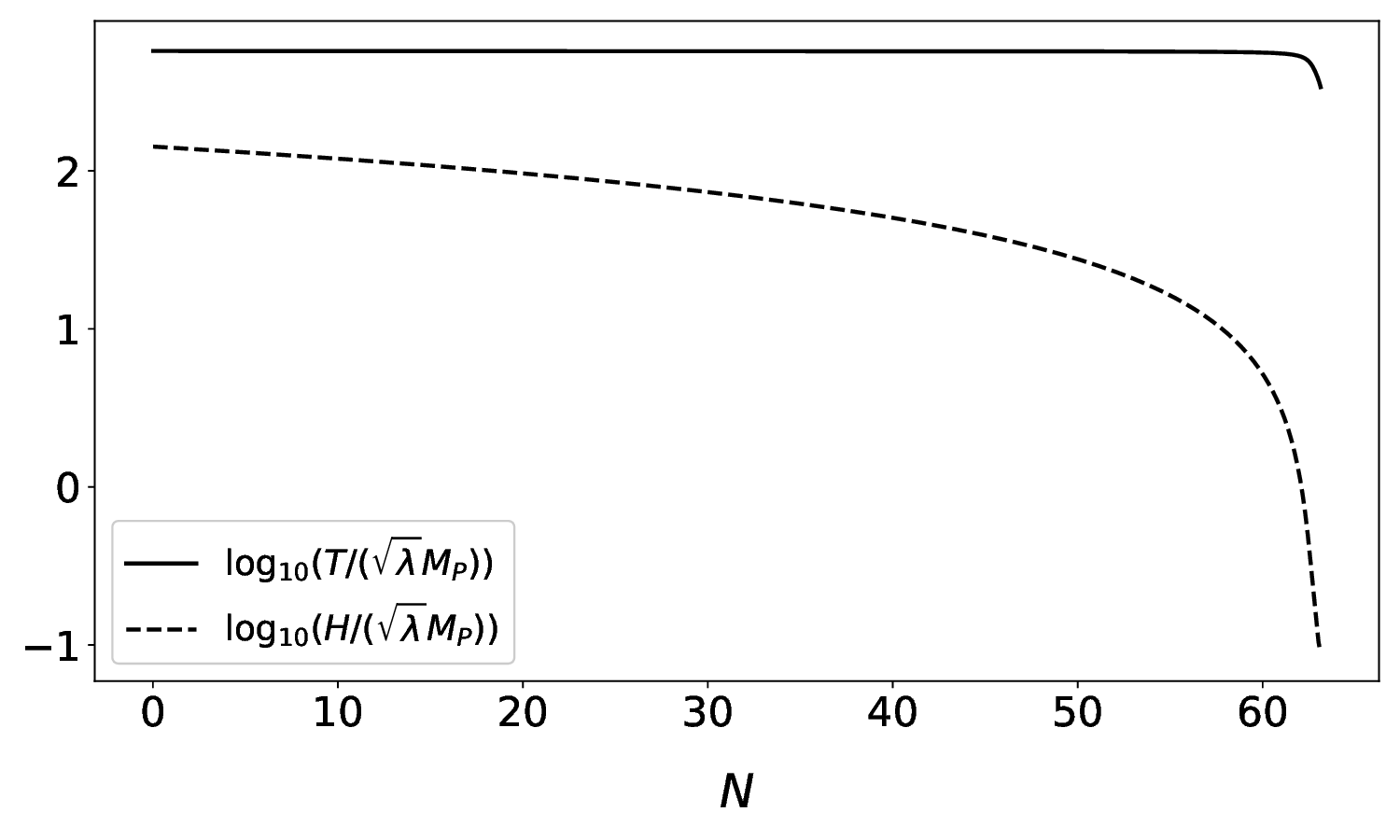}
\caption{Comparison of the temperature $T$ and the Hubble rate $H$ in the weak dissipative regime of quartic warm inflation.}
\label{fig-2}
\end{figure*}

Inflation ends when the slow-roll parameter $\eta=\frac{12}{\phi^2}M_P^2$, as defined in \eqref{slow-roll}, reaches the value of unity. In the weak regime, this condition corresponds to the inflaton field $\phi_f$ at the end of inflation, which is given by $\phi_f= 2\sqrt 3 M_P$.

Using equation \eqref{fried1a}, the number of efolds is estimated as
\begin{equation}
\label{eefold}
N=\int_{t_i}^{t_f} H dt= \frac{1}{M_P^2}\int_{\phi_f}^{\phi_i}\frac{V}{V'}d\phi = \frac{({\phi_i}^2-{\phi_f}^2)}{8M_P^2}
\end{equation}
with approximations $\rho\ll\dot\phi^2$ and $\dot\phi^2\ll \lambda\phi^4$ during slowroll.
Considering $N=60$, and without neglecting $\phi_f$ in equation \eqref{eefold}, we obtain the initial value of the inflaton field as
\begin{equation}\label{phi_i}
\phi_i=22.1811 \,M_P.
\end{equation}
Furthermore, equation \eqref{phidot} provides the initial value of $\dot\phi_i$, which is given by
\begin{equation}
\label{phidot1}
\dot\phi_i=-25.6125 \, \lambda M_P^2.
\end{equation}

In the weak regime of warm inflation, the power spectrum amplitude described by equation \eqref{Power}, simplifies to
\begin{equation}
\label{power-weak}
{A_R}^{1/2} \simeq \left(\frac{3H^3}{2\pi V^\prime}\right)\left(\frac{T}{H}\right)^{1/2},
\end{equation}
that takes the form
\begin{equation}
\label{power-weak}
{A_R}^{1/2} \simeq \frac{N}{\pi}\left(\frac{\lambda\sqrt a}{6\sqrt C_*}\right)^{1/3},
\end{equation}
upon using equations \eqref{T-weak} and \eqref{fried1a} with the quartic potential \eqref{potential}.

\begin{figure*}
\centering
\includegraphics[width=0.8\linewidth]{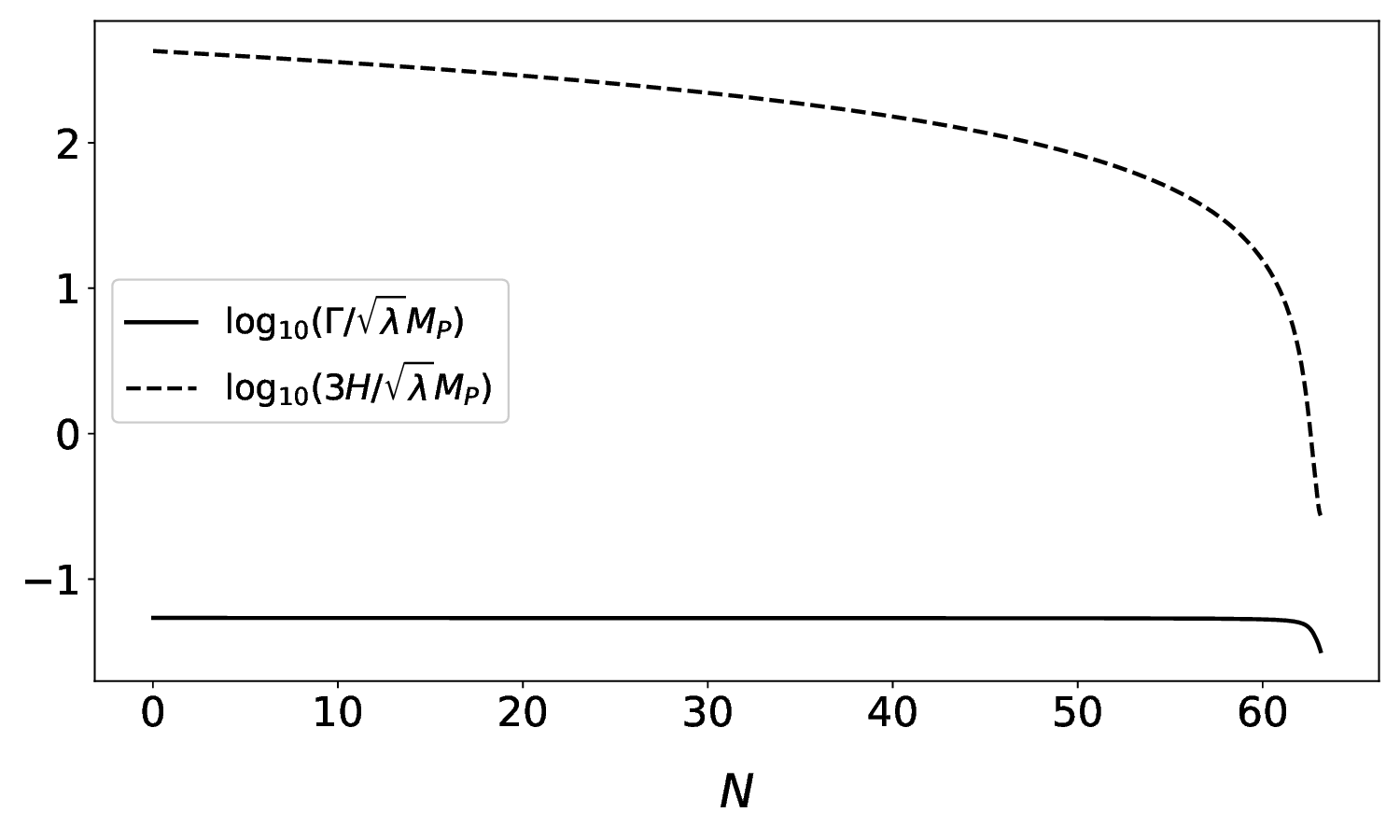}
\caption{Comparison between the evolutions of the dissipative coefficient $\Gamma$ and Hubble expansion rate $H$ in the weak dissipative regime.}
\label{fig-3}
\end{figure*}

In the weak dissipative regime, with $R\ll 1$, equations \eqref{ns1} and \eqref{r1} reduce to
\begin{equation}
n_s = 1 - 2\epsilon + 2\sigma
\label{ns2}
\end{equation}
and
\begin{equation}
r = 16 \frac{H}{T} \epsilon.
\label{r2}
\end{equation}
These equations can be combined to yield
\begin{equation}
r = \frac{4}{5^5} \sqrt{\frac{C_*}{a}} (1-n_s).
\label{comb1}
\end{equation}
Based on this relation, Figure \ref{fig-a} illustrates that the value $a=9.4421 \times 10^{-5}$ is well within the bounds of the observed Planck data \cite{Planck-2013,Planck-2015}.

\subsection{The strong dissipative regime}

\label{strong-reg}
In the strong regime, interaction between the inflaton field and the surrounding thermal bath is so strong that energy transfer occurs at a much faster rate than the expansion of the Universe; in other words, $\Gamma\gg 3H$ or $R\gg 1$.  Using \eqref{dissipation}, equation \eqref{temperature} therefore modifies to
\begin{equation}
\label{T-strong}
T=\left(\frac{1}{4C_*a}\right)^{1/5}\left(\frac{V'^2}{H}\right)^{1/5},
\end{equation}
and equation \eqref{KG1a} leads to
\begin{equation}
\label{KG2}
\dot\phi=-\frac{V'(\phi)}{\Gamma}=-\frac{\lambda \phi^3}{\Gamma}
\end{equation}
for the quartic potential \eqref{potential}.

In this regime, the condition which is to be satisfied at the end of inflation follows from violation of the the slow roll condition \eqref{condition}, which may be taken as $\eta=R$. Noting from \eqref{slow-roll} that $\eta=\frac{12}{\phi^2}M_P^2$,   this condition by equations \eqref{dissipation} and \eqref{T-strong} give the value of the inflaton field
\begin{equation}\label{phi-f}
\phi_f=\frac{\lambda^{1/4}}{a}\left(\frac{6^7 C_*}{2}\right)^{1/4}\ M_P
\end{equation}
at the end of inflation.

The number of efolds can be obtained in strong regime as
\begin{eqnarray}
N=\int_{t_i}^{t_f} H dt= \frac{1}{M_P^2}\int_{\phi_f}^{\phi_i}\frac{V}{V'}Rd\phi\nonumber\\
{=\frac{5}{4}\left(\frac{1}{4\lambda C_*}\right)^{1/5}\left(\frac{a}{2\sqrt 3 M_P}\right)^{4/5}\left(\phi_i^{4/5}-\phi_f^{4/5}\right)},
\label{eefold2}
\end{eqnarray}
that clearly depends on both $a$ and $\lambda$, unlike the expression \eqref{eefold} in the weak regime. This equation can give the initial value of $\phi_i$ for $N=60$ upon using equation \eqref{phi-f} when the values of $a$ and $\lambda$ are supplied.

By substituting equations \eqref{dissipation} and \eqref{T-strong} into \eqref{KG2}, we obtain the initial value of $\dot\phi$ as
\begin{equation}
\label{KG2a}
\dot\phi_i=-\left(\frac{2C_*\lambda^{7/2}}{\sqrt 3 a^4}\right)^{1/5}\left(\frac{\phi_i}{M_P}\right)^{11/5}M_P^2,
\end{equation}
which can be evaluated once $\phi_i$ is obtained from \eqref{eefold2} upon supplying the values of $a$ and $\lambda$.

\begin{figure*}
\centering
  \includegraphics[width=0.8\linewidth]{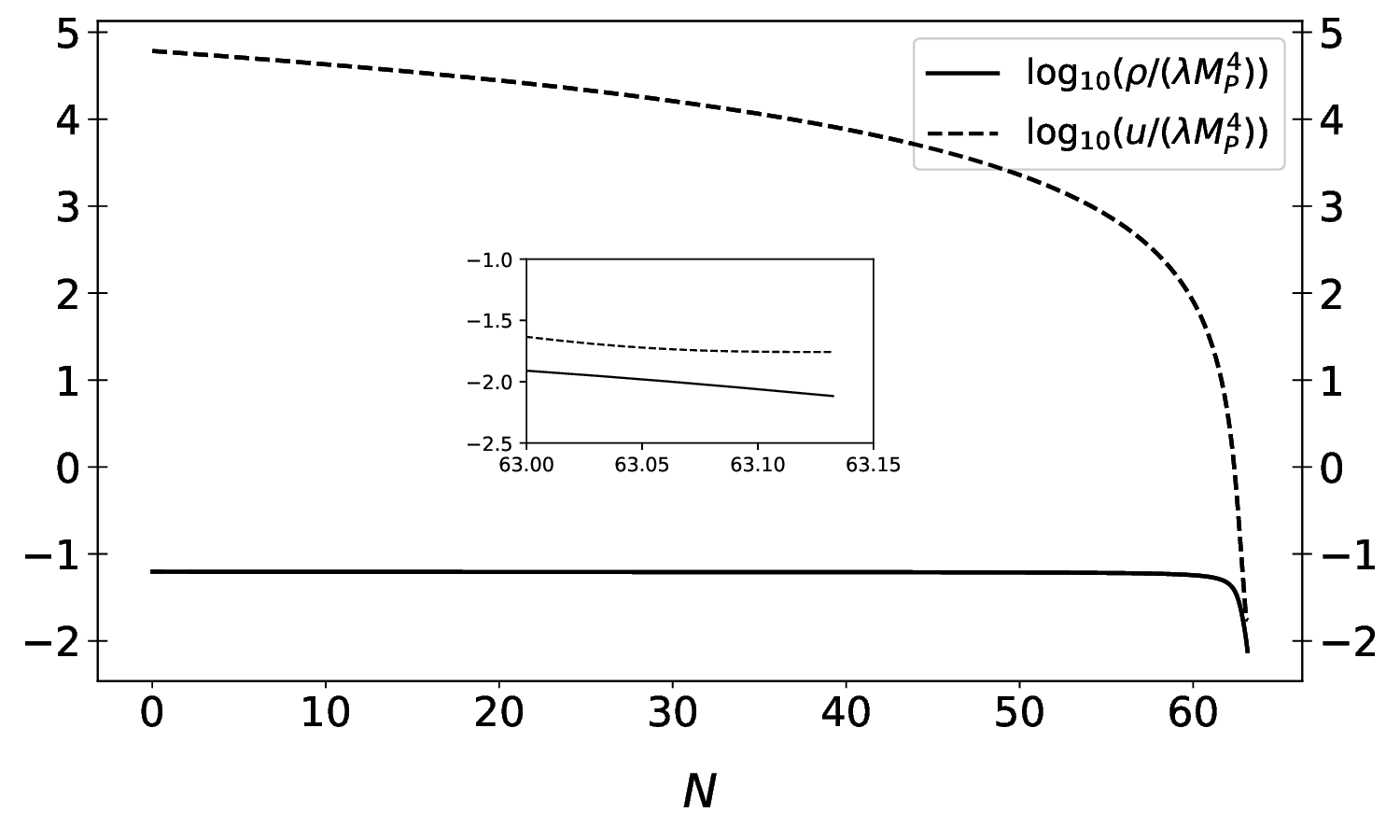}
\caption{Comparison of the evolutions of inflaton energy density $u$ and radiation energy density $\rho$ in the weak dissipative regime of quartic warm inflation. The inset illustrates their behaviours after the end of inflation.}
\label{fig-4}
\end{figure*}

We shall use the initaial conditions $\phi_i$ and $\dot\phi_i$ to solve the dynamical equations of warm inflation in the strong regime.

The amplitude of power spectrum in the strong regime of warm inflation is greatly influenced by dissipation and thermal effects. Thus equation \eqref{Power} takes the form as
\begin{equation}
\label{power-strong}
{A_R}^{1/2} \simeq \left(\frac{3H^3}{2\pi V^\prime}\right)\left(\frac{T}{H}\right)^{1/2} R^{5/4}.
\end{equation}
For the quartic model described by \eqref{potential}, and by using \eqref{fried1a} and \eqref{T-strong}, we obtain
\begin{equation}
\label{power-strong}
{A_R }^{1/2}\simeq \frac{2}{\pi} \frac{1}{(9C_*)^{1/8}}
\left(\frac{\lambda N^3}{125}\right)^{3/8}.
\end{equation}

As noted earlier, we shall use ${A_R}^{1/2}=10^{-5}$ \cite{Planck-2013,Planck-2015}. Thus, with $N=60$, we find that $\lambda=7.6775\times 10^{-16}$.

Since the interaction strength $\lambda$ in the quartic potential \eqref{potential} should be a universal quantity, we shall take this value of $\lambda$ irrespective of the nature of regime of warm inflation.

\begin{figure*}
\includegraphics[scale=0.3]{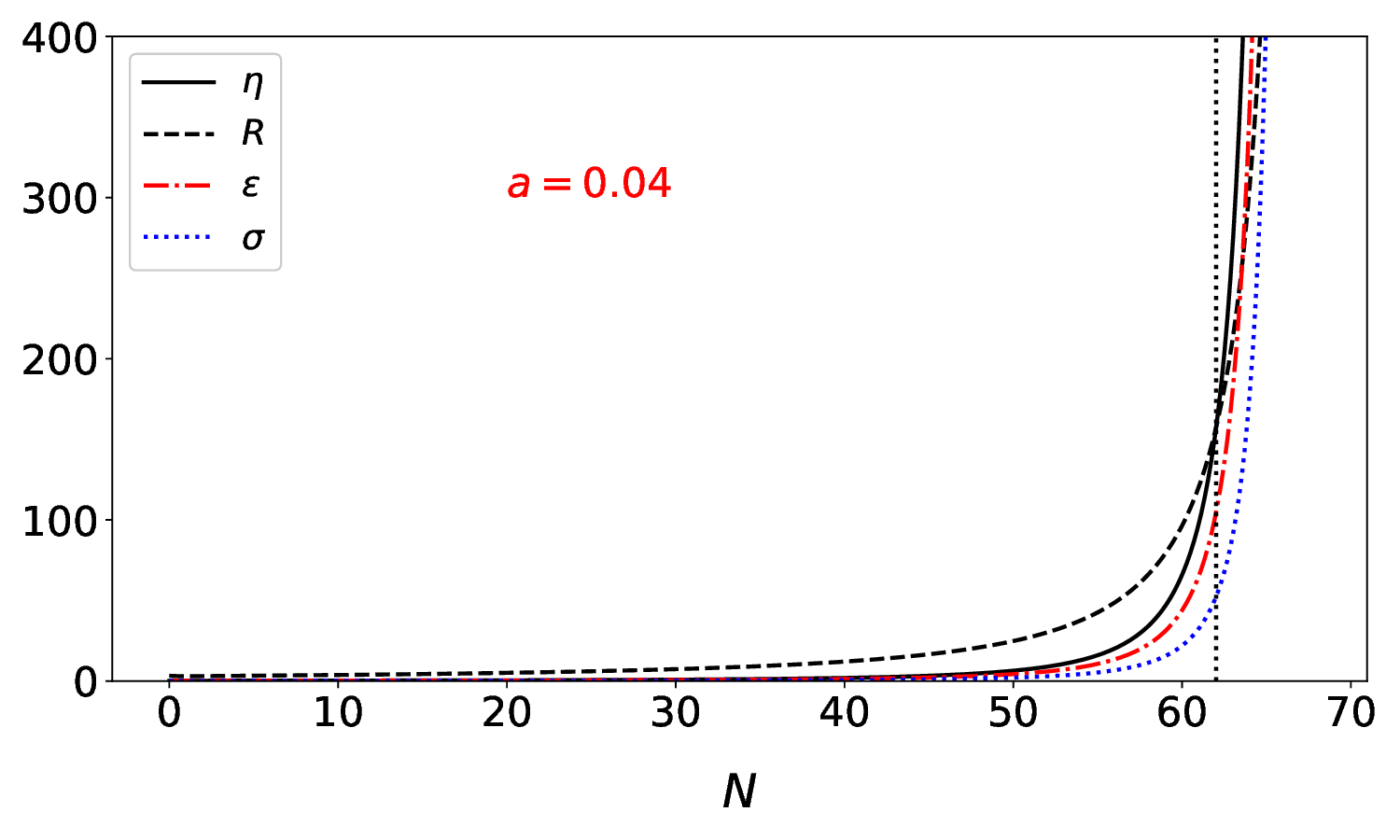}
\includegraphics[scale=0.3]{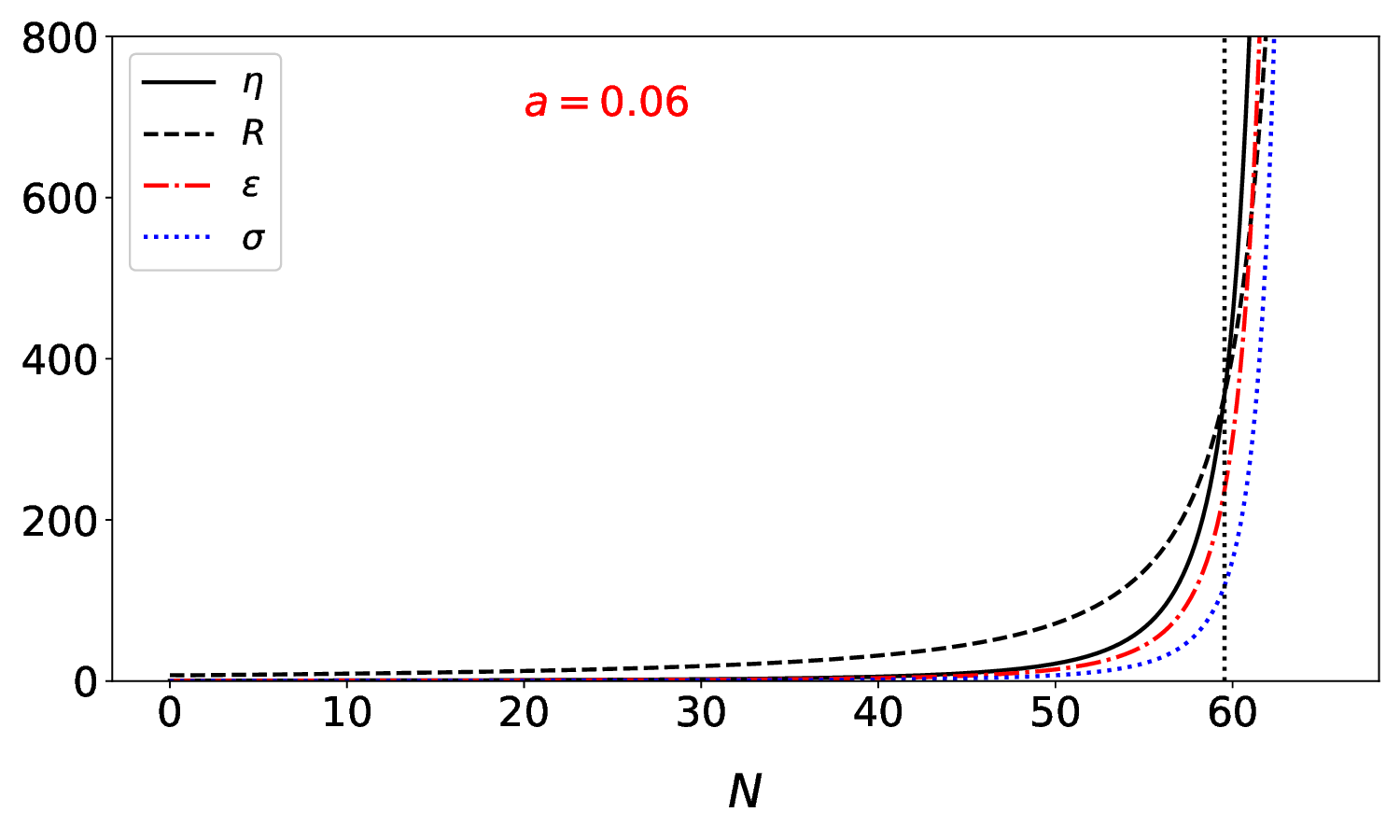}
\includegraphics[scale=0.3]{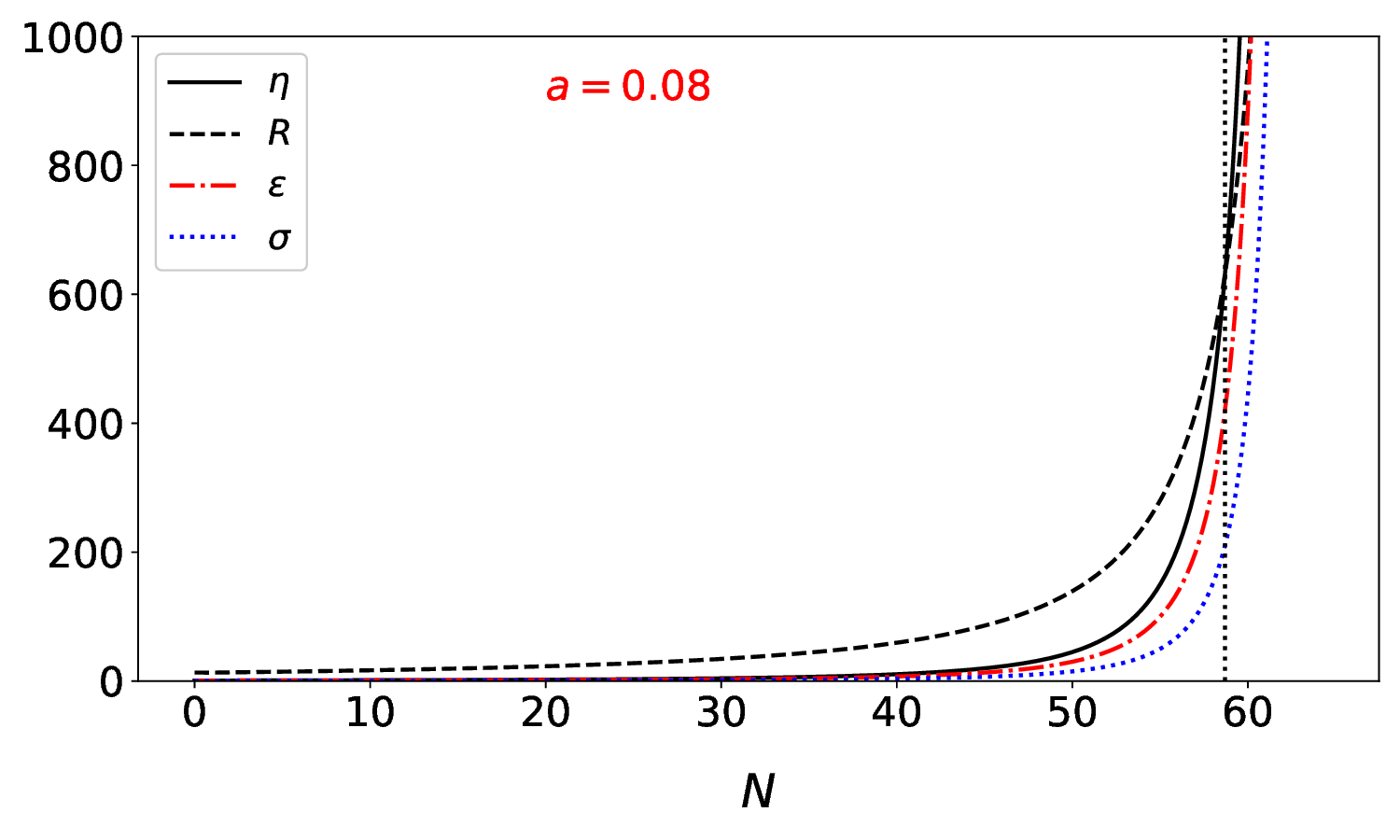}
\includegraphics[scale=0.3]{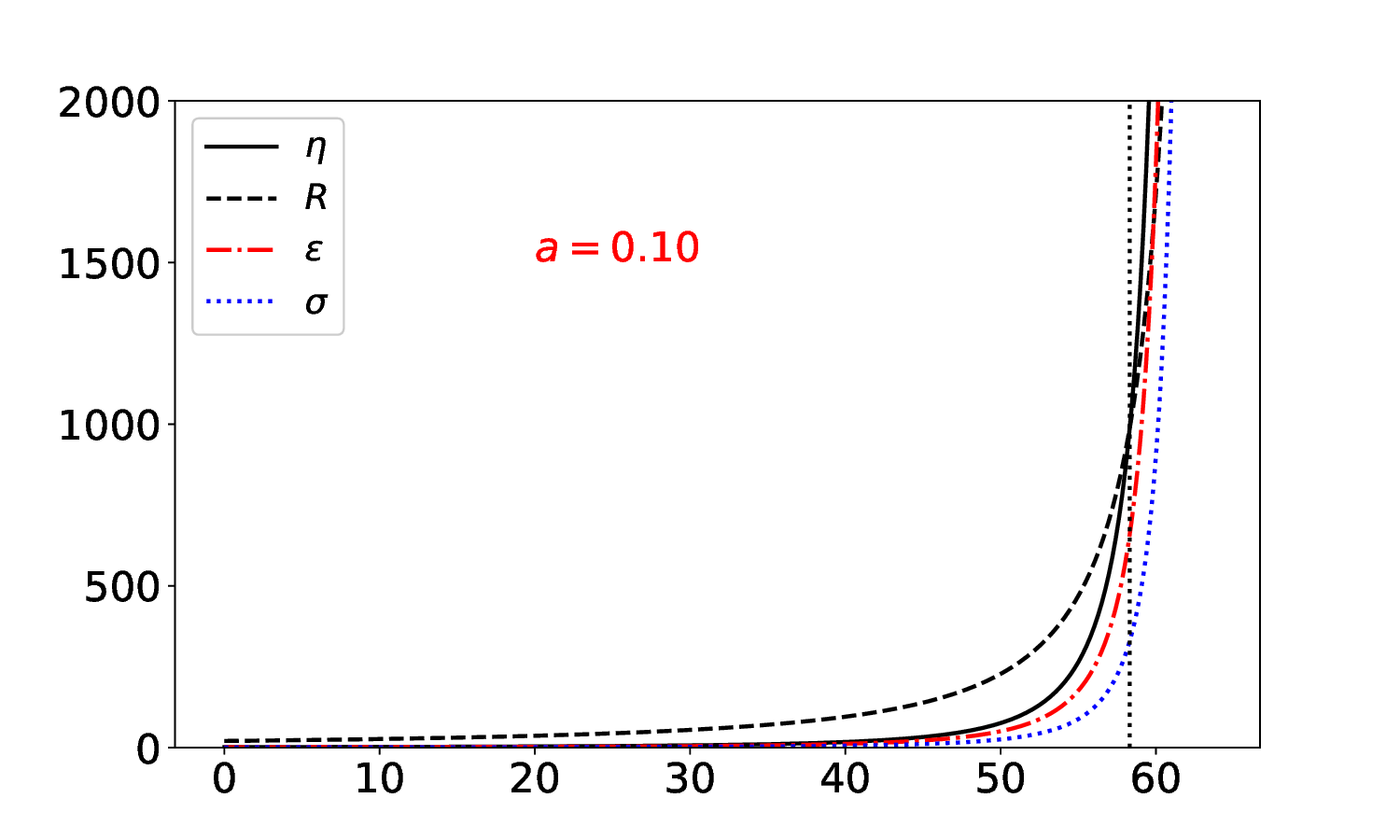}
\caption{Evolution of the slow roll parameters $\epsilon$, $\eta$, $\sigma$ and the dissipation ratio $R$ in the strong dissipative regime for $a=0.04, 0.06, 0.08$ and $0.10$. The vertical lines indicate $N=62.0160,59.5437,58.7000$ and $58.3135$, where inflation ends when $\eta=R$.}
\label{fig-5}
\end{figure*}

In the strong dissipative regime, with $R\gg 1$, equations \eqref{ns1} and \eqref{r1} assume the forms
\begin{equation}
n_s = 1 + \frac{1}{7R} ( -9\epsilon - 3\eta + 18\sigma )
\label{ns3}
\end{equation}
and
\begin{equation}
r = \frac{H}{T} \frac{16\epsilon}{R^{5/2}}.
\label{r3}
\end{equation}
By combining these equations, we obtain
\begin{equation}
r = 8.5 \times 10^{-9} \frac{\pi^{10/3}}{a^4} (1-n_s).
\label{comb2}
\end{equation}
Based on this relation, Figures \ref{fig-b}, \ref{fig-c}, \ref{fig-d}, and \ref{fig-e} illustrate that the values $a=0.04, 0.06, 0.08, 0.10$ are well within the bounds of the observed Planck data \cite{Planck-2013,Planck-2015}.

\section{Evolutionary Dynamics of Warm Inflation}
\label{sec4}

In the previous sections, we analysed the inflationary phase with a dissipative coefficient linear in temperature by applying the standard approxiamtions for warm inflation with slow roll conditions. 
Such analytical approximations are expected to yield a quasi-static evolution in the initial phase of inflation.  As time progresses, and particularly in the later phase of inflation, these approximations may not fully capture the nonlinear dynamics inherent in the coupled evolution of the inflaton and radiation fields. 
This necessitates the {\em exact} solution of the complete set of strongly nonlinear coupled dynamical equations, without any approximation or neglect of term, using numerical integration.

Thus, in order to explore the exact dynamical evolution of warm inflation of the Universe, we shall numerically integrate the rate equations for the inflaton energy density $u$ and the radiation energy density $\rho$ given by \eqref{KG} and \eqref{reheat} along with the Friedmann equation \eqref{fried1}, that may be written in convenient forms as
\begin{equation}
\label{udotnew}
\dot{u}=-3\sqrt{\frac{u+\rho}{3M_P^2}}\,\,\left(2u-\frac{1}{2}\lambda \phi^4\right)-\Gamma \dot \phi^2,
\end{equation}

\begin{equation}
\label{rhodotnew}
\dot{\rho}=-4\sqrt{\frac{u+\rho}{3M_P^2}}\,\,\rho+\Gamma \dot \phi^2,
\end{equation}

\begin{equation}
\label{phidotnew}
\dot{\phi}^2=2u-\frac{1}{2}\lambda \phi^4,
\end{equation}
upon using the quartic potential \eqref{potential} in \eqref{inflaton} and \eqref{pressure}. The above equations have to be supplemented with $\Gamma = aT=a\left(\frac{\rho}{C_*}\right)^{1/4}$.

The strongly nonlinear and coupled nature of the differential equations \eqref{udotnew}, \eqref{rhodotnew}, and \eqref{phidotnew} precludes any exact analytical solutions and they necessitate numerical integration to achieve accurate estimates for the time evolution of the inflaton and radiation energy densities.

\begin{figure*}
\includegraphics[scale=0.3]{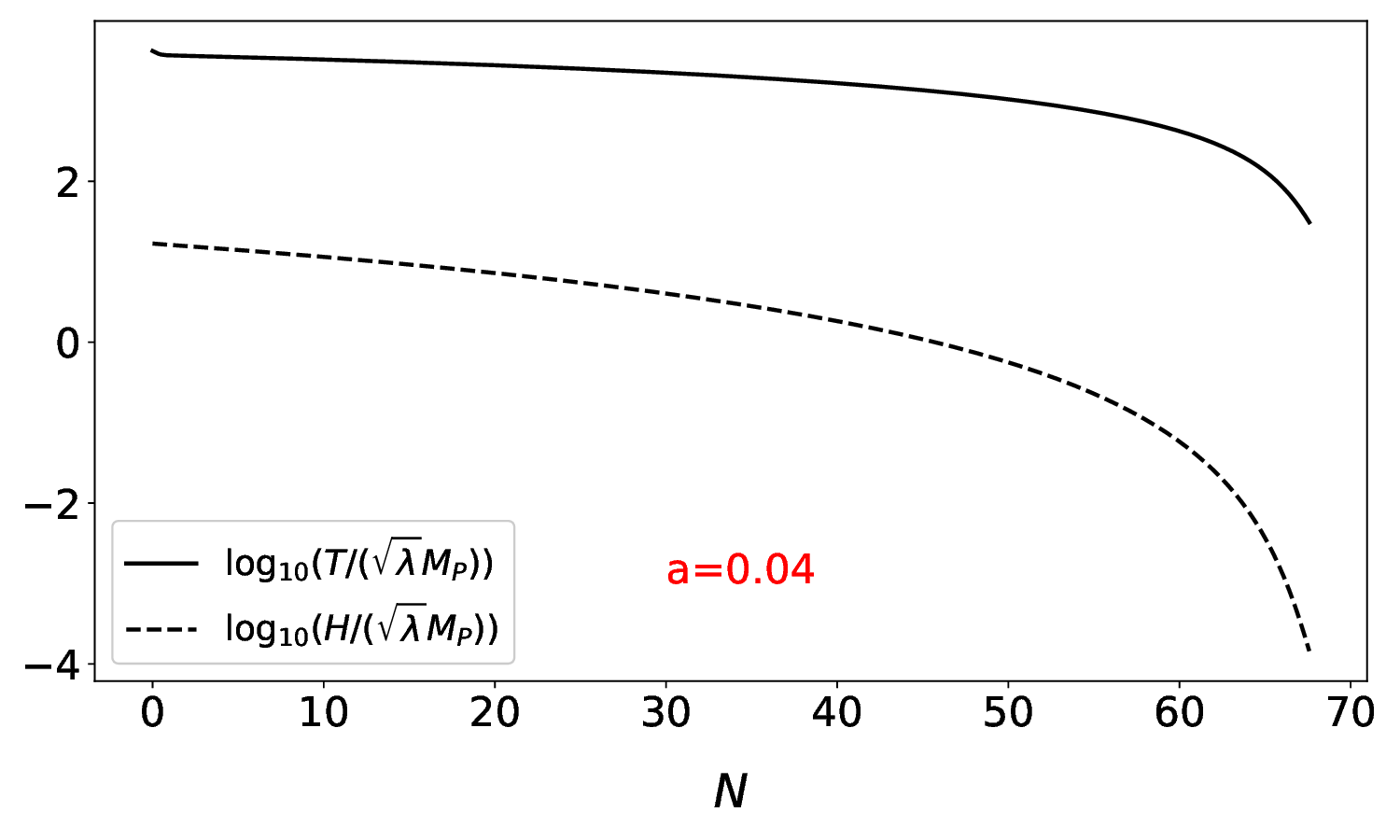}
\includegraphics[scale=0.3]{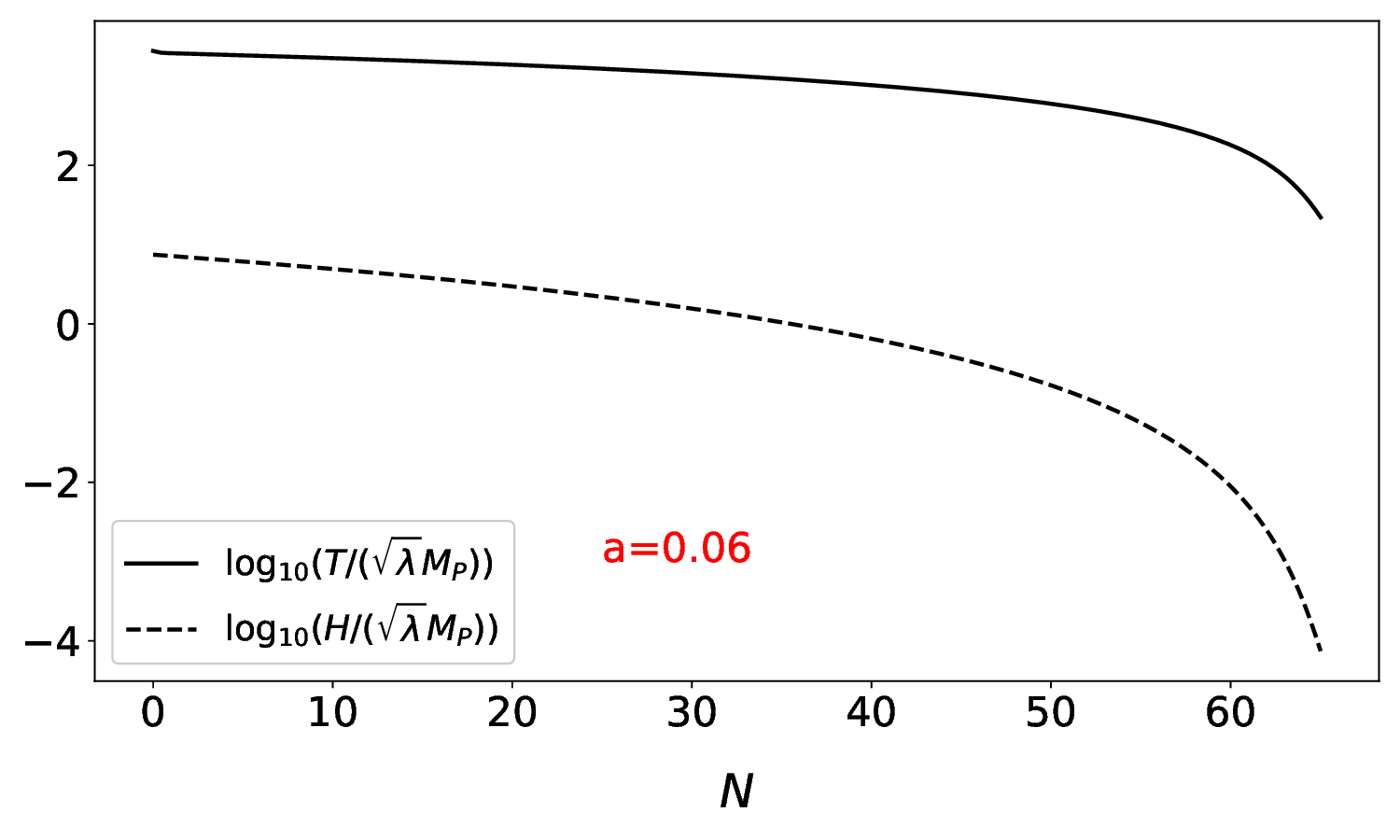}
\includegraphics[scale=0.3]{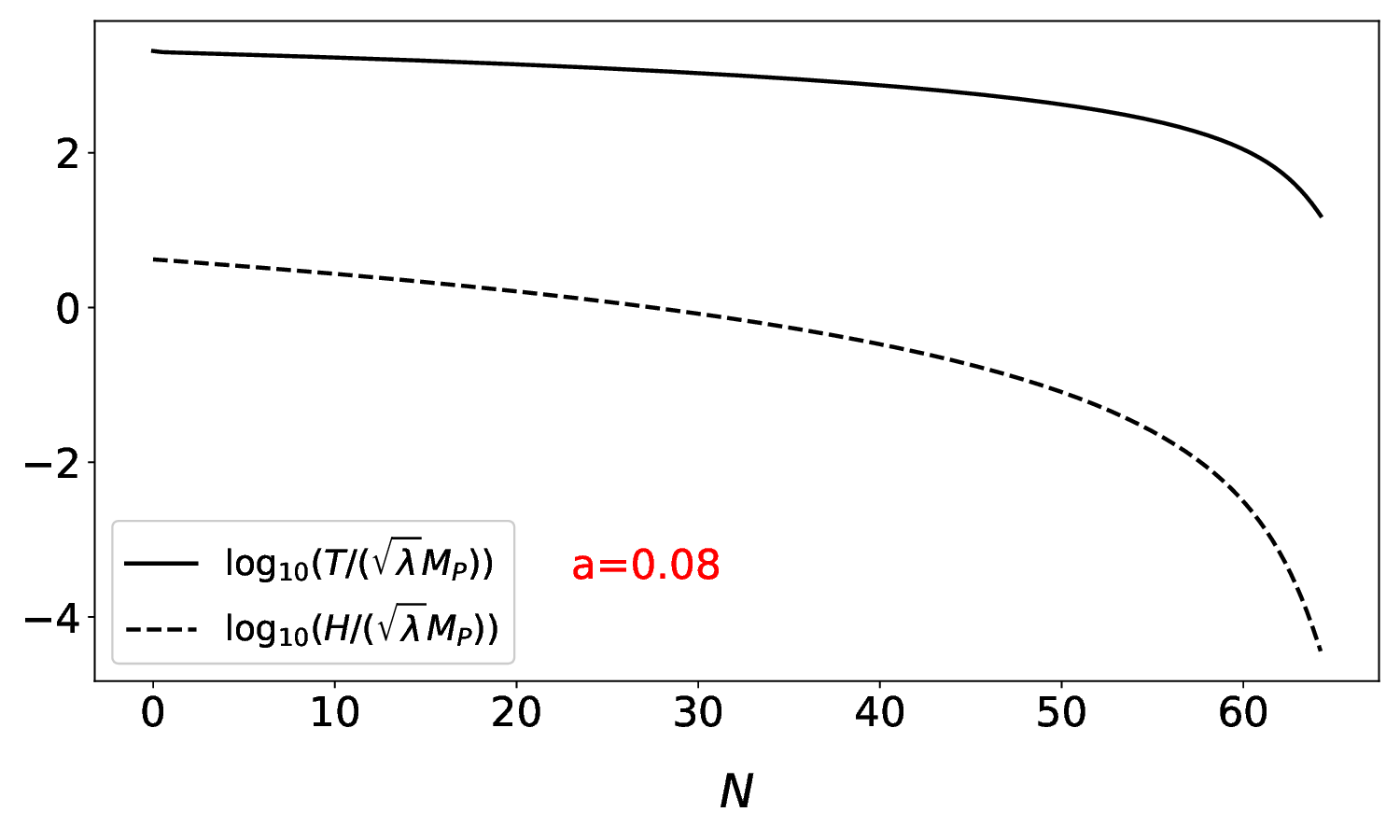}
\includegraphics[scale=0.3]{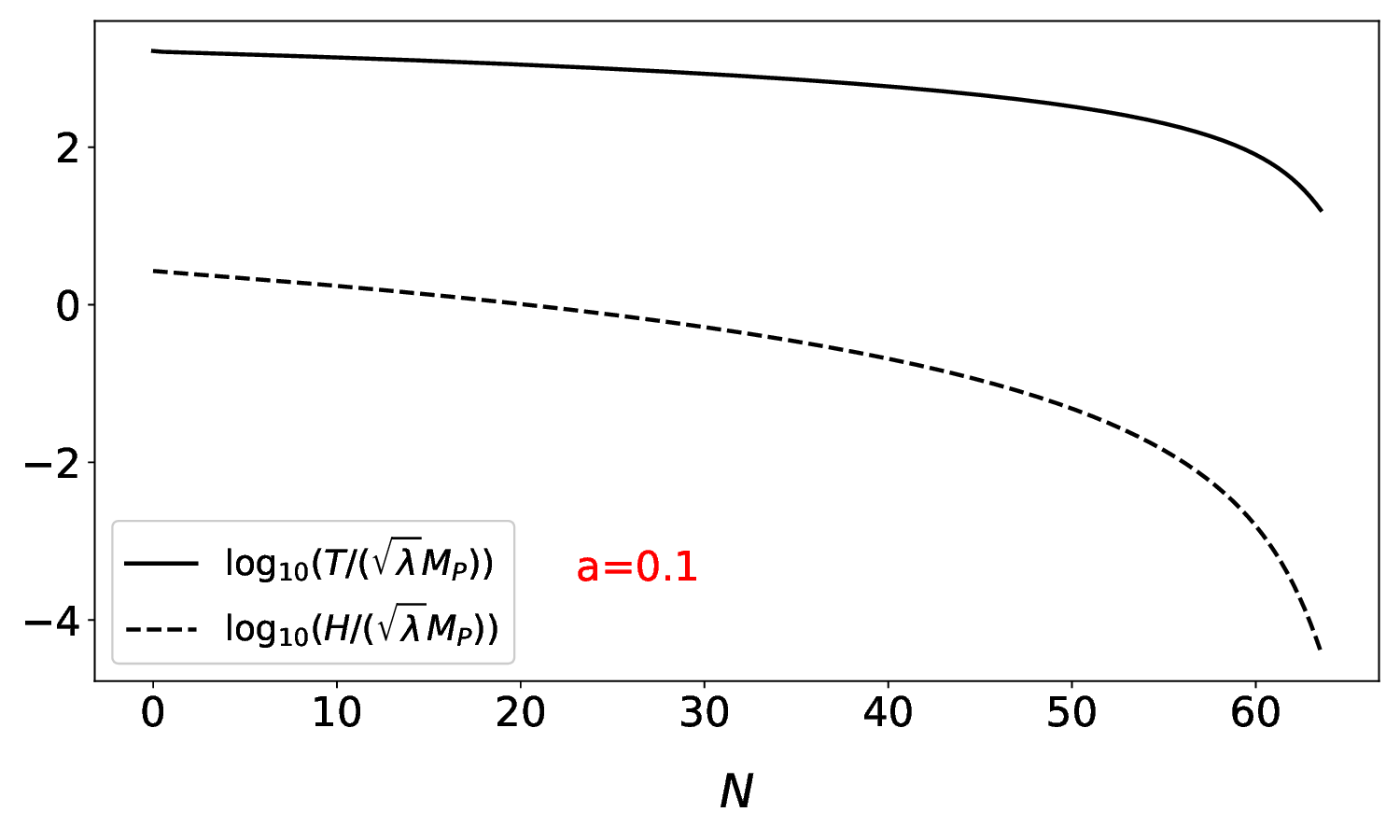}
\caption{Comparison of the temperature $T$ and the Hubble expansion rate $H$ in the strong dissipative regime of quartic warm inflation for $a=0.04, 0.06, 0.08$ and $0.10$.}
\label{fig-6}
\end{figure*}

\begin{figure*}
\includegraphics[scale=0.3]{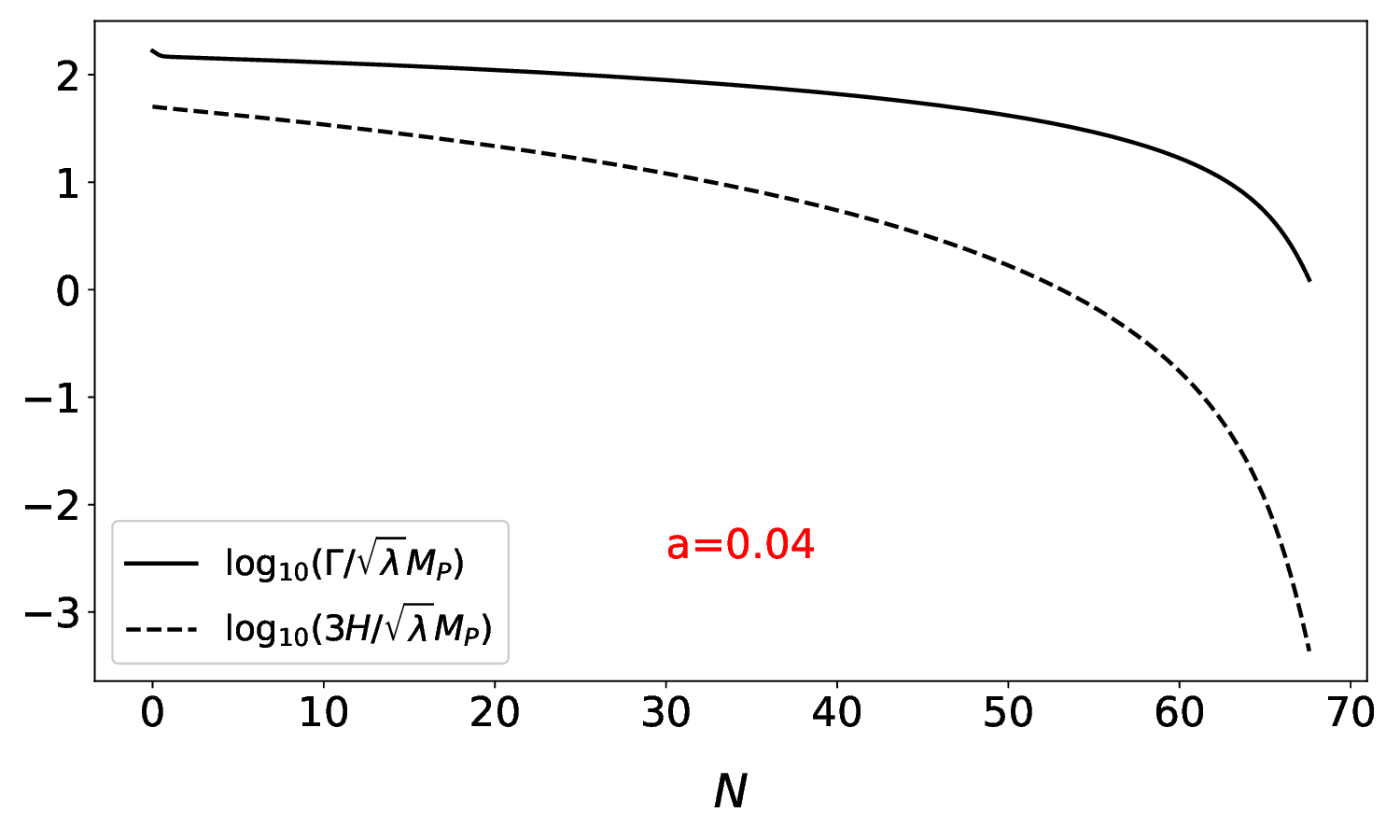}
\includegraphics[scale=0.3]{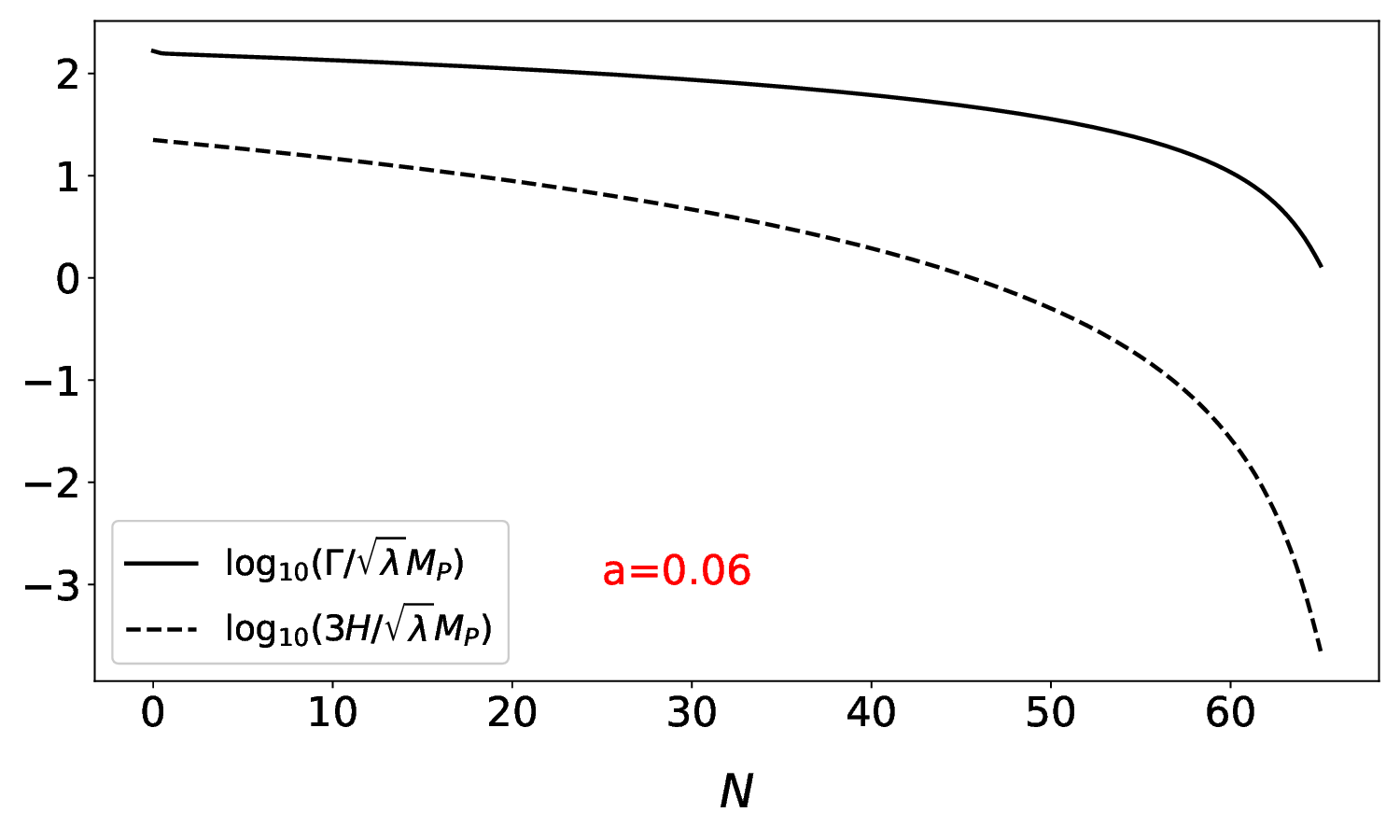}
\includegraphics[scale=0.3]{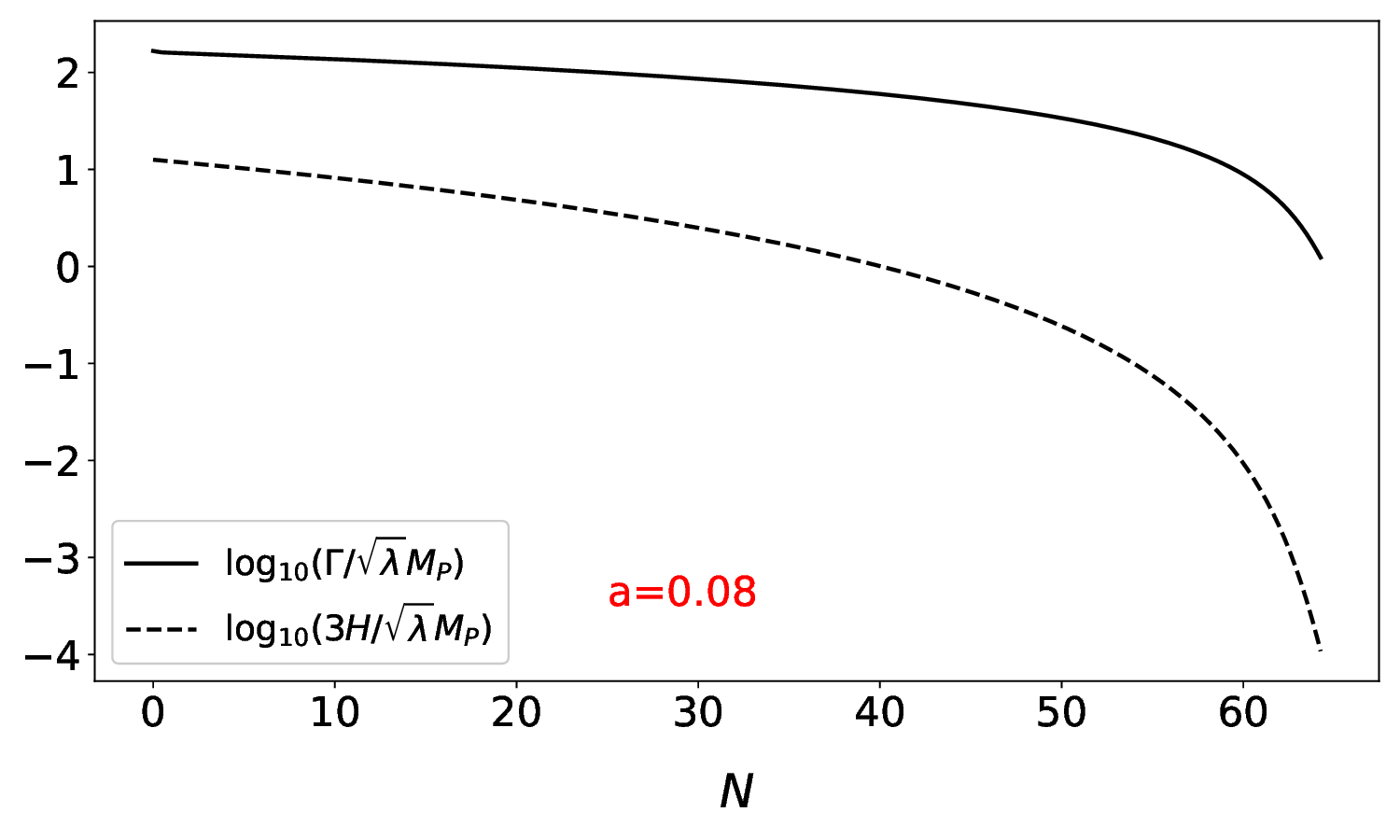}
\includegraphics[scale=0.3]{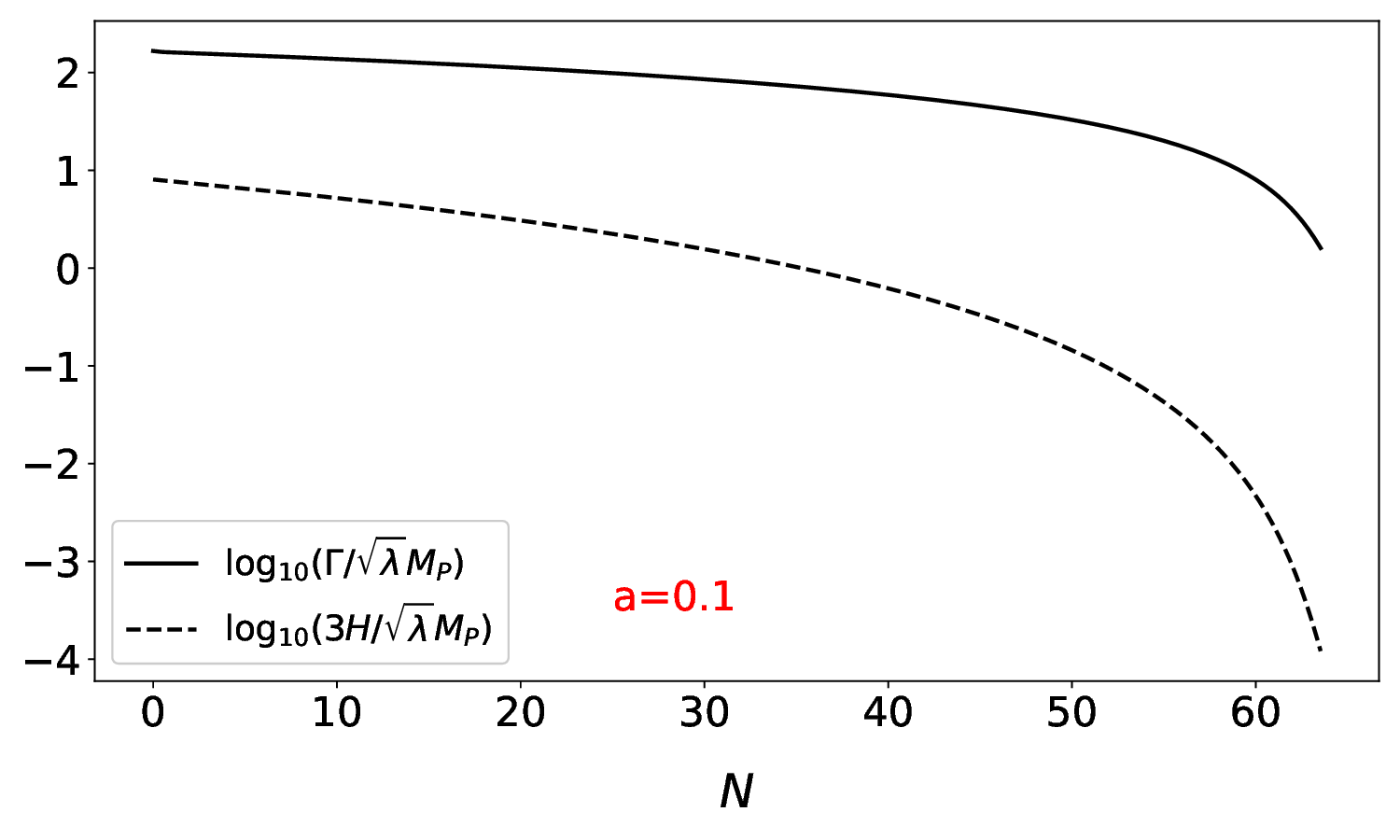}
\caption{Evolution of the dissipative coefficient $\Gamma$ and the Hubble expansion rate $H$ in the strong dissipative regime for $a=0.04, 0.06, 0.08$ and $0.10$.}
\label{fig-7}
\end{figure*}

To simplify and streamline the process of numerical integration, we rescale the variables as
\begin{equation}
\label{rescale}
\tau=\sqrt \lambda M_P t, \ \ \tilde{\phi}=\frac{\phi}{M_P}, \ \ \tilde{u}=\frac{u}{\lambda M_P^4}, \ \ \tilde{\rho}=\frac{\rho}{\lambda M_P^4}.
\end{equation}
As a result, the dimensionalized versions of the coupled differential equations \eqref{udotnew}, \eqref{rhodotnew} and \eqref{phidotnew} are
\begin{equation}
\label{u-tilde}
\frac{d\tilde{u}}{d\tau}=-3\sqrt{\frac{\tilde{u}+\tilde{\rho}}{3}}\,\,\left(2\tilde{u}-\frac{1}{2}\tilde{\phi}^4\right)-\tilde a {\tilde{\rho}}^{1/4}\left(\frac{d\tilde{\phi}}{d\tau}\right)^2,
\end{equation}
\begin{equation}
\label{rho-tilde}
\frac{d\tilde{\rho}}{d\tau}=-4\sqrt{\frac{\tilde{u}+\tilde{\rho}}{3}}\,\,\tilde{\rho}+\tilde a {\tilde{\rho}}^{1/4}\left(\frac{d\tilde{\phi}}{d\tau}\right)^2,
\end{equation}
and
\begin{equation}
\label{phi-tilde}
\left(\frac{d\tilde{\phi}}{d\tau}\right)^2=2\tilde{u}-\frac{1}{2}\tilde{\phi}^4,
\end{equation}
where $\tilde a=\frac{a}{(\lambda C_*)^{1/4}}$.

In order to obtain the dynamical evolution of warm inflation without any approximation, we shall numerically integrate equations \eqref{u-tilde}, \eqref{rho-tilde} and \eqref{phi-tilde} {\em without dropping any terms}. In carrying out the numerical integration, we shall use appropriate initial conditions at the initial time $t_i$ previously obtained in Section \eqref{basics}.

\begin{figure*}
\includegraphics[scale=0.3]{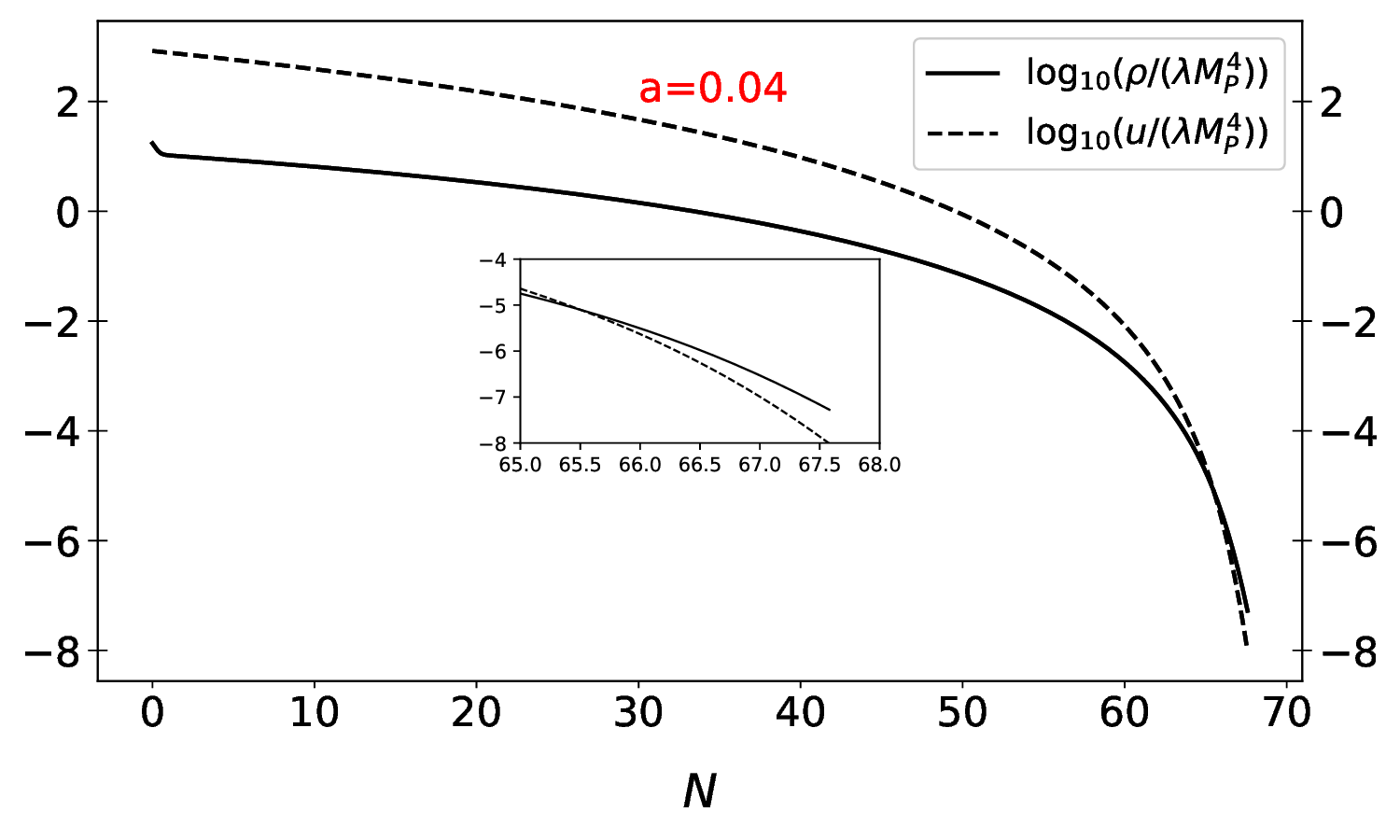}
\includegraphics[scale=0.3]{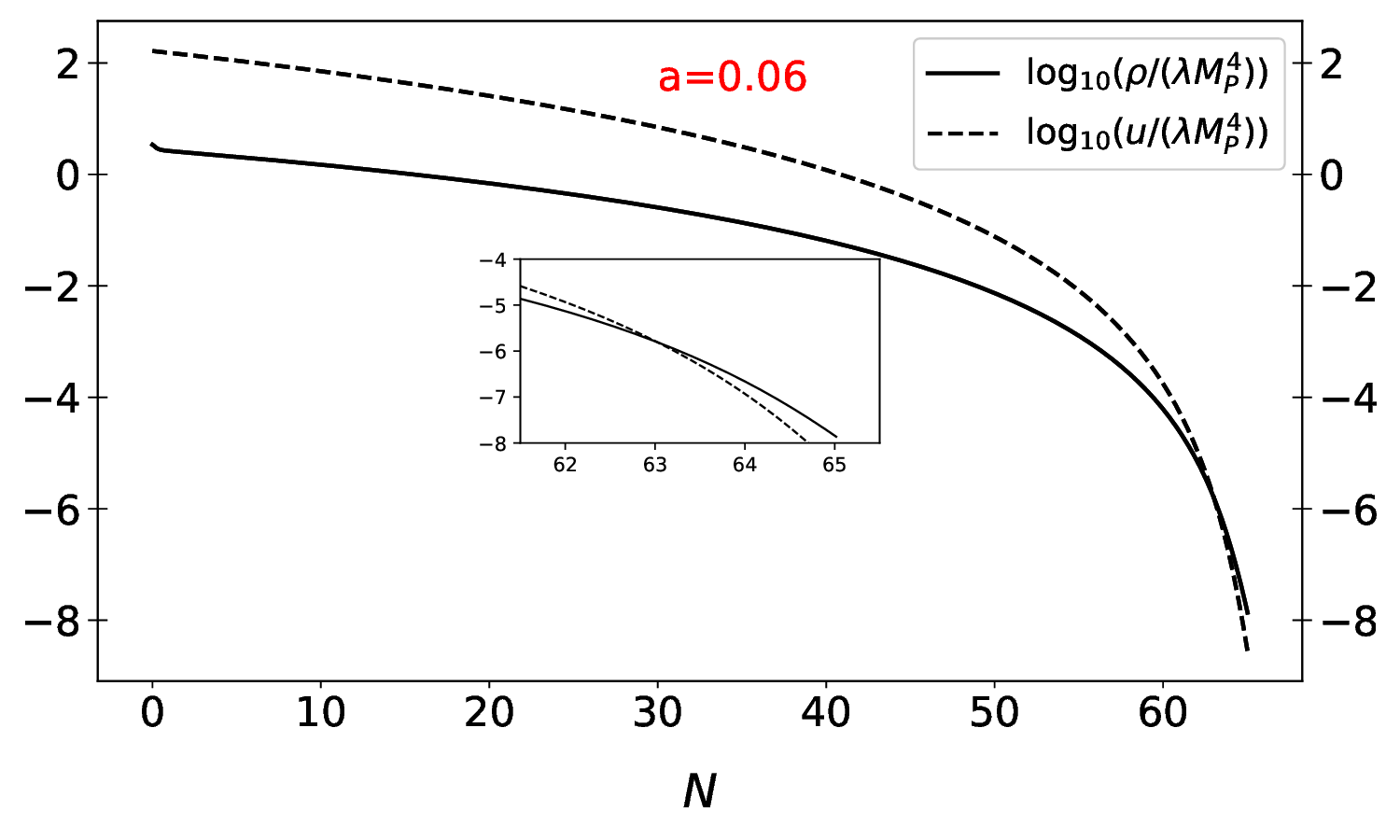}
\includegraphics[scale=0.3]{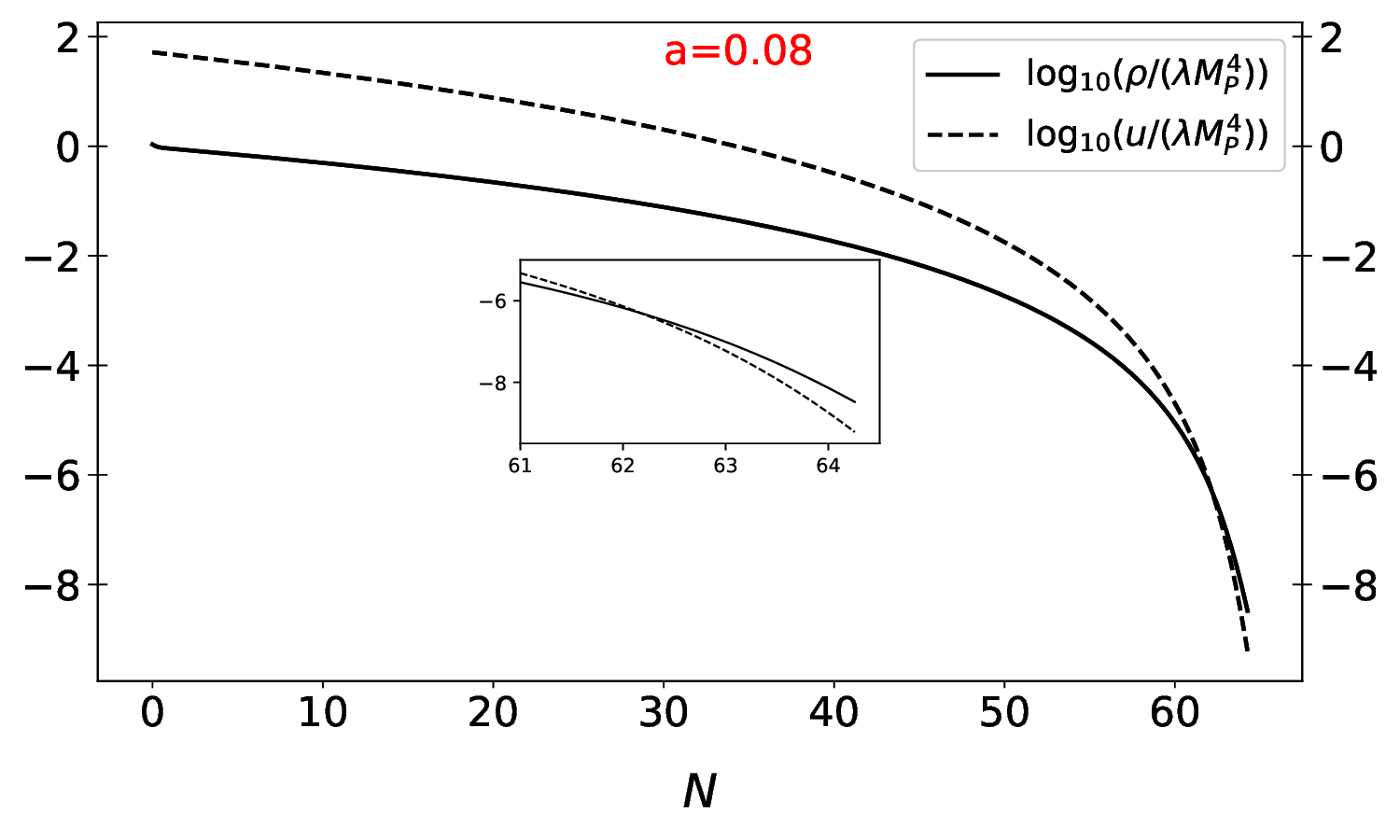}
\includegraphics[scale=0.3]{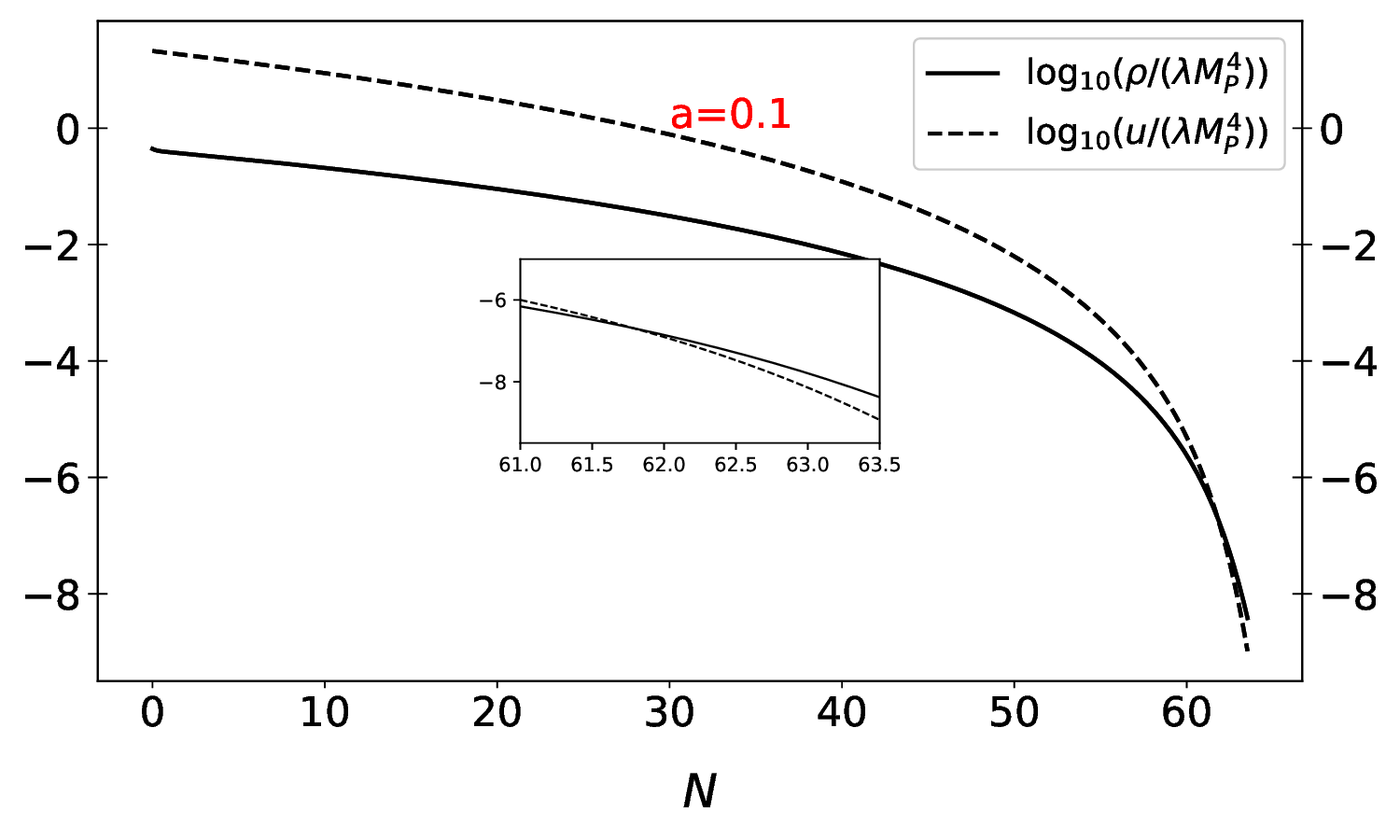}
\caption{Comparison of the evolutions of inflaton energy density $u$ and radiation energy density $\rho$ in the strong dissipative regime of quartic warm inflation for $a=0.04, 0.06, 0.08$ and $0.10$. The insets illustrate that the inflaton-radiation crossover takes place slightly beyond the end of inflation.}
\label{fig-8}
\end{figure*}

\subsection{Evolution in the weak regime}
The amplitude of the power spectrum \eqref{power-weak} gives $a=9.4421 \times 10^{-5}$ for  ${A_R}^{1/2}=10^{-5}$ and $\lambda=7.6775\times 10^{-16}$. Using these parameter values, we integrate numerically the coupled set of nonlinear differential equations \eqref{u-tilde}, \eqref{rho-tilde} and \eqref{phi-tilde} with the initial conditions for the weak regime.

Figure \eqref{fig-1} shows the evolutions of the slow roll parameters $\epsilon$, $\eta$ and  $\sigma$. We clearly see that all three slow roll parameters remain close to zero in most part of the inflationary period. It is only towards the end of inflation that they begin to rise to higher values. Upon identifying the end of inflation with $\eta=1$, we see that the inflation ends at $N=61.5986$. In our model of $\Gamma=aT$, the slow roll parameter $\beta$ remains zero.

Figure \eqref{fig-2} shows the plots for the temperature $T$ and the Hubble expansion rate $H$ throughout the weak regime of warm inflation. It is clear that the condition $T>H$ is satisfied throughout this regime.

In figure \eqref{fig-3}, we compare the time evolutions of the dissipative coefficient $\Gamma$ and Hubble expansion rate $H$. The weak dissipation condition $\Gamma\ll H$ for the weak regime is seen to be satisfied in the course of warm inflation.

Figure \eqref{fig-4} compares the evolution of the inflaton energy density $u$ with the radiation energy density $\rho$. As shown in the figure, the inflaton energy density remains dominant over radiation throughout the inflationary period. The inset highlights that this dominance persists even after inflation ends at $N=61.5986$. This behavior arises because the dissipation coefficient $\Gamma$ is too small ($\Gamma\ll H$), indicating that the weak regime is not suitable for the Universe for transitioning into a radiation-dominated phase.

\begin{table*}[h!]
\centering
\caption{Values of different quantities at the end of inflation (identified with $\eta\approx R$, denoted by $_f$ as subscript) in the strong dissipative regime, for  different values of the parameter $a$.}
\begin{tabular}{|c|c|c|c|c|c|c|c|}
\hline
$a$	     & $N_f$	         & $\tilde u_f$ 		      & $\tilde \rho_f$             & $\eta_f$	    & $R_f$ 		     & $\tilde T_f$                       &          $T_f$ in GeV \\ \hline
$0.04$	 & $62.0160$     & $1.4537\times 10^{-3}$ & $4.5456\times 10^{-4}$    & $157.4493$	& $157.4472$	 & 297.8223
         & $2.0053\times 10^{13}$      \\ \hline
$0.06$   & $59.5437$     & $2.8586\times 10^{-4}$ & $8.9955\times 10^{-5}$    & $354.9560$	& $354.9550$	 & 198.6397     		 & $1.3375\times 10^{13}$      \\ \hline
$0.08$   & $58.7000$     & $9.0313\times 10^{-5}$ & $2.8482\times 10^{-5}$    & 631.4413	& 631.4403	     & 149.0059    		 & $1.0033\times 10^{13}$      \\ \hline
$0.10$   & $58.3135$     & $1.1670\times 10^{-5}$ & $3.6965\times 10^{-5}$    & 986.9577	& 986.9362	     & 119.2137   		     & $8.0268\times 10^{12}$      \\ \hline

\end{tabular}
\label{strong}
\end{table*}

\subsection{Evolution in the strong regime}

In the strong dissipative regime, the amplitude of  power spectrum given by equation \eqref{power-strong} depends only on the interaction strength $\lambda$, that gives $\lambda=7.6775\times 10^{-16}$ with the observed value $A_R = 10^{-5}$.
For the parameter $a$, we consider three possible values, $a=4\times 10^{-2},6\times 10^{-2}$, $8\times 10^{-2}$ and $ 10^{-1}$ \cite{Panotopoulos2015}, to perform the numerical integration of the coupled nonlinear differential equations \eqref{u-tilde}, \eqref{rho-tilde}, and \eqref{phi-tilde}, with initial conditions corresponding to the strong regime.

Figure \eqref{fig-5} illustrates evolution of the slow roll parameters $\epsilon$, $\eta$ and $\sigma$ in the strong dissipative regime for four parameter values $a=4\times 10^{-2},6\times 10^{-2}$, $8\times 10^{-2}$ and $ 10^{-1}$. In all cases, the slow roll parameters maintain their closeness to zero in most part of the inflationary period. Towards the end of inflation, these parameters begin to rise. We have identified the end of inflation when $\eta$ grows up to the value of $R$. This is shown as vertical lines in the plots. As before, the slow roll parameter $\beta=0$ in the considered model $\Gamma= aT$.

Table \eqref{strong} presents the values of various quantities obtained from our numerical simulations for the three different values of $a$. The table clearly indicates that larger values of $a$ necessitate a greater number of e-folds to complete inflation. Additionally, the condition $\eta = R$, marking the end of inflation, is achieved at higher values of $\eta$ as $a$ increases.

Figure \eqref{fig-6} displays the temperature $T$ and the Hubble expansion rate $H$ in the strong regime of warm inflation. It is clear that the condition of $T>H$ for warm inflation holds good in the inflationary regime in all cases of $a$.

In figure \eqref{fig-7}, we make comparisons between the dissipative coefficient $\Gamma$ and Hubble expansion rate $H$ for the three values of $a=0.04, 0.06, 0.08$ and $0.10$. We see that the condition $\Gamma\gg 3H$ for the strong regime is satisfied throughout the course of warm inflation for all cases of $a$.

Figure \eqref{fig-8} illustrates the evolution inflaton energy density $u$ and radiation energy density $\rho$ in the strong dissipative regime of warm inflation for the three values of $a$.

\begin{table*}[h!]
\centering
\caption{Values of different quantities at the radition-inflaton crossover (identified with $u\approx \rho$, denoted by $_c$ as subscript) in the strong dissipative regime, for different values of the parameter $a$.}
\begin{tabular}{|c|c|c|c|c|c|c|}
\hline
$a$	     & $N_c$	         & $\tilde u_c$ 		          & $\tilde\rho_c$                & $R_c$ 		 & $\tilde T_c$ & $T_c$ in GeV \\ \hline
$0.04$	 & 65.5054       & $7.8639\times 10^{-6}$     & $7.8638\times 10^{-6}$    	& $628.9763$ & $108.0109$  & $7.2725\times 10^{12}$      \\ \hline
$0.06$   & 63.6247       & $1.5616\times 10^{-6}$     & $1.5616\times 10^{-6}$    	& 1413.3150	 & 72.1034     & $4.8548\times 10^{12}$     \\ \hline
0.08     & 62.1781       & $4.9505\times 10^{-7}$     & $4.9505\times 10^{-7}$    	& 2511.3683	 & 54.1032     & $3.6428\times 10^{12}$      \\ \hline
0.10     & 61.7902       & $2.0295\times 10^{-7}$     & $2.0295\times 10^{-7}$    	& 3923.1580	 & 43.2920     & $2.9149\times 10^{12}$      \\ \hline
\end{tabular}
\label{strong-1}
\end{table*}

Table \eqref{strong-1} presents the values of several quantities obtained from our numerical computations for the three different values of $a$. In comparison with Table \eqref{strong}, Table \eqref{strong-1} shows that the inflaton-radiation crossover ($u = \rho$) occurs approximately three e-folds after inflation ends. For all values of $a$, the temperature of the Universe drops to about $10^{12}$ GeV at this crossover, following the end of inflation at around $10^{13}$ GeV, as indicated in Table \eqref{strong-1}. Subsequently, the Universe transitions into the radiation-dominated phase ($\rho > u$), as illustrated in the insets of Figure \eqref{fig-8}.

Importantly, the Universe smoothly transitions to a radiation-dominated phase soon after inflation ends in the strong regime. It takes about four e-folds to make this transition as seen from a comparison between Tables \ref{strong} and \ref{strong-1} for each value of the parameter $a$. The temperature of the Universe at this point of is found to be about $10^{12}$ GeV.  This indicates that the the Universe gracefully exits to a hot, radiation-dominated phase.

\section{Discussion and Conclusion}
\label{Discussion}

In this work, we went beyond the usual approximations in warm inflation and our main focus was upon the exact nonlinear dynamical nature of warm inflation with the quartic potential for the inflaton field. The observational data for the amplitude of the power spectrum was used to fix the coupling constant in the quartic model.  Upon modeling the dissipation coefficient as a linear function of temperature, we obtained the rate equations for the inflaton energy density and the radiation energy density, both driving the Friedmann equation. This led to a set of three coupled strongly nonlinear differential equations, not amenable to solve by analytical means. With appropriate initial conditions derived from the slow roll conditions in consistency with Planck data, we {\em exactly} solved the nonlinear differential equations governing the dynamics of inflaton and radiation,  numerically in the weak and strong regimes of warm inflation, {\em without dropping any terms}.

Our numerical integration of the three nonlinear coupled differential equations yielded the exact dynamical evolution of several important quantities. Consequently, we compared the time evolution of the inflaton energy density and the radiation energy density in both weak and strong regimes of warm inflation (Figures \eqref{fig-4} and \eqref{fig-8}). Within the framework of our model, we found that the Universe does not go into a radiation dominated phase even after inflation ends in the weak regime, which is an obvious consequence of the dissipative coefficient being too small in this regime. This conclusion follows from the fact that the Hubble parameter dominates over the dissipative coefficient throughout the weak regime (Figure \eqref{fig-3}) of warm inflation. On the other hand, the dissipative coefficient always dominates over the Hubble parameter in the strong regime (Figure \eqref{fig-7}). Dissipation being too strong, the Universe can enter into a radiation dominated phase in the strong regime, which is found to occur soon after inflation is over (Figure \eqref{fig-8}).

Despite these contrastive features between the two regimes, the temperature of the Universe always dominates over the Hubble expansion rate in both regimes (Figures \eqref{fig-2} and \eqref{fig-6}), an important characteristic of warm inflation.

Our numerical simulation in the strong regime shows that the temperature of the Universe becomes $\sim 10^{13}$ GeV at the end of inflation that cools to a temperature of $\sim 10^{12}$ GeV at the point of radiation-inflaton equality. Subsequently, the Universe enters into to a hot, radiation dominated phase.

Thus, our consideration of the exact nonlinear dynamics of warm inflation with the dissipative coefficient $\Gamma\propto {T}$ and a quartic inflaton potential $V\propto\phi^4$ shows that the Universe enters into a radiation dominated phase in the strong regime, thus solving the graceful exit problem.

\section*{Acknowledgements}
Bhargabi Saha is supported by a Research Fellowship from the Ministry of Education, Government of India. The authors thank IIT Guwahati for computational support.


\end{document}